\documentclass[aps,pra,twocolumn,superscriptaddress]{revtex4-2}

\usepackage{amsthm}
\usepackage{amsmath,bm}
\usepackage{amssymb}
\usepackage{amsfonts}
\usepackage{graphicx}
\usepackage{txfonts}
\usepackage{xcolor}
\usepackage{float}
\usepackage{braket}

\usepackage[squaren]{SIunits}

\usepackage[colorlinks=true,linkcolor=blue,citecolor=blue,urlcolor=blue]{hyperref}

\usepackage{bbold}

\newcommand{\avg}[1]{\langle#1\rangle}		

\newcommand{\op}[1]{\hat{#1}}				

\newcommand{\ie}{\textit{i.e.} }

\begin{document}

\title{Quantum gravitational decoherence of a mechanical oscillator from spacetime fluctuations}

\author{Sandro Donadi}
\email{s.donadi@qub.ac.uk}
\affiliation{Centre for Quantum Materials and Technologies, School of Mathematics and Physics, Queens University, Belfast BT7 1NN, United Kingdom}
\affiliation{Istituto Nazionale di Fisica Nucleare, Trieste Section, Via Valerio 2, 34127 Trieste, Italy}

\author{Matteo Fadel}
\email{fadelm@phys.ethz.ch}
\affiliation{Department of Physics, ETH Z\"{u}rich, 8093 Z\"{u}rich, Switzerland}

\begin{abstract}
We consider the scenario of a fluctuating spacetime due to a deformed commutation relation with a fluctuating deformation parameter, or to a fluctuating metric tensor. By computing the resulting dynamics and averaging over these fluctuations, we find that a system experiences a decoherence in the momentum basis. We studied the predictions of the model for a free particle and an harmonic oscillator. Using experimental data taken from a mechanical oscillator prepared in quantum states of motion, we put a bound on the free parameters of the considered model. In addition, we comment on how these measurements can also provide bounds to other phenomenological quantum gravity models, such as the length scale for nonlocal dynamics.
\end{abstract}

\maketitle

Quantum gravity theories, as well as considerations from Black hole physics, suggest a fundamental structure of space-time characterized by a minimal observable length scale \cite{veneziano_stringy_1986,amati_can_1989,konishi_minimum_1990,maggiore_generalized_1993,garay_quantum_1995,SCARDIGLI199939,Adler99,Hossenfelder13}. One possible way in which this could arise is from a modification of the commutation relation between position and momentum in quantum mechanics \cite{maggiore93,Kempf97,fadelPRD22}. 
An extensively studied example of deformed commutator is
\begin{equation}\label{eq:ModComm}
    [\op{X},\op{P}] = i\hbar \left( 1 + \frac{\beta\ell_{P}^{2}}{\hbar^{2}} \op{P}^2 \right) \;,
\end{equation}
where $\ell_{P} = \sqrt{\hbar G/c^3}$ is the Planck length and $\beta$ the unitless deformation parameter, typically considered to be $\beta \sim 1$. 
Equation~\eqref{eq:ModComm} results in the generalized uncertainty principle (GUP)
\begin{equation}\label{eq:gup}
\Delta\op{X} \Delta\op{P} \geq \dfrac{\hbar}{2} \left( 1+\frac{\beta\ell_{P}^{2}}{\hbar^{2}} \Delta\op{P}^2 \right) \;,
\end{equation}
which indeed predicts a minimal observable length scale. In fact, by minimizing over $\Delta\op{P}$ yields $\Delta \op{X} \geq \ell_P \sqrt{\beta}$, namely that the localization in position cannot be arbitrary small.

While the minimal length scale is often considered to be a constant, it was recently suggested the possibility of having a foamy spacetime in which this quantity fluctuates. This could be effectively described by a deformed commutator Eq.~\eqref{eq:ModComm} with time-dependent $\beta$ \cite{Petru21}. Interestingly, a minimal length scale that fluctuates stochastically can result in a universal decoherence mechanism for quantum systems.

Investigating the structure of spacetime at the Planck scale is therefore of great interest for shedding light on the interplay between gravity and quantum mechanics, and for improving our understanding of the most fundamental laws of Nature. Unfortunately, however, such a scale cannot be accessed by direct observation, as it would require experiments at energies which are many orders of magnitude beyond our current capabilities. Nevertheless, it has been noticed that deformed commutators can give rise to nontrivial effects also at low energies, meaning that they might be observed experimentally using available technologies. 

In particular, significant interest was raised by the modifications to the physics of a mechanical harmonic oscillator resulting from deformed commutators of the type of Eq.~\eqref{eq:ModComm} with constant $\beta$. For example, there is a correction to the energy spectrum and to the oscillator wave functions \cite{Kempf97,Das16}, as well as an effective nonlinearity (\ie the presence of non-quadratic terms in the Hamiltonian) which results in an amplitude-dependent oscillation frequency and in a third-harmonic generation \cite{AliPRD11,Pedram12,Bosso17}. Even if of small magnitude, these effects are expected to appear also at the energy scales that are accessible by state-of-the-art experiments, and could thus be probed by precision measurements. 
In addition, since the momentum is proportional to the mass, it has been suggested that the effects of a deformed commutator could even be amplified by considering the center-of-mass degrees of freedom of massive systems \cite{Bosso17,IgorGUP12}.

For the above reasons, considerable effort has been put in probing deformed commutators using massive mechanical oscillators. Recent experiments in this direction performed measurements on oscillators with masses ranging from $\unit{10^{-11}}{kg}$ to $\unit{10^{3}}{kg}$, to put bounds on the deformation parameter $\beta$ \cite{bawaj15,Tobar19,marinNat13}.
Despite these remarkable results, however, it is crucial to highlight that all these experiments were performed in the classical regime, and not in the quantum one. It is thus debated what conclusions can be drawn from these sort of measurements, as relating them to a Planck scale modification of quantum mechanics relies on the additional assumption that a deformed commutation relation also results in deformed Poisson brackets \cite{BenczikPRD02,Scardigli15,SCARDIGLI2017242,kumarNC20}.

Using quantum systems to investigate deformed commutators is therefore of significant interest, as interpreting the results does not rely on this additional assumption. Moreover, as we will show, by looking at the time evolution of nonclassical states and at their decoherence dynamics, a quantum system could give access to exploring whether the deformation parameter, and thus the minimal length scale, fluctuates.

In this work, we first derive the dynamics of a quantum system subject to a deformed commutation relation Eq.~\eqref{eq:ModComm} with fluctuating $\beta$, including non-Markovian effects. 
In particular, we focus on the dynamics of an harmonic oscillator and show how energy eigenstates and their superposition evolve in the presence of stochastic fluctuations of this parameter.
Then, we consider experiments where a $\unit{16}{\mu g}$ mechanical resonator is prepared in such nonclassical states of motion and investigate their time evolution. 
From the measured energy relaxation and decoherence rates, we are able to put a bound on the fluctuations of $\beta$. 
Finally, we show how a precise measurement of the ground state variances of the mechanical resonator can allow us to bound the time-averaged value of $\beta$. We conclude by showing how these same techniques can be used to constrain the length scale associated to quantum gravity induced nonlocal dynamics \cite{BelenchiaPRL} and another model proposed by Breuer \textit{et al.} predicting decoherence in momentum due to fluctuations of the spacetime metric \cite{breuer2009metric}.

\vspace{2mm}
\textbf{Modified dynamics.--} We want to derive the effect of the modified algebra Eq.~\eqref{eq:ModComm} on the dynamics of a quantum system. For this, it is convenient to introduce the transformation
\begin{equation}
\hat{X} = \hat{x} \;, \qquad \hat{P} = \left(1+\frac{\beta\ell_{P}^{2}}{\hbar^{2}}\hat{p}^{2}\right)\hat{p} \;, \label{relXxPp} 
\end{equation}   
such that the lowercase position and momentum operators satisfy the canonical commutation relation $[\op{x},\op{p}]=i\hbar$ to order $\mathcal{O}(\beta^2)$. Expressing the system's Hamiltonian in terms of these new operators results in the modified dynamics 
\begin{equation}
i\hbar\frac{d}{dt}|\psi\rangle=\left[\hat{H} + \hat{H}_{\beta} \right]|\psi\rangle \;, \label{ModDyn}
\end{equation}
where $\hat{H}=\hat{K}+V(\hat{x})$, $\hat{K}=\hat{p}^{2}/2m$ is the kinetic energy,  
\begin{equation} \label{Hb}
\hat{H}_\beta=4a_{P} \beta \hat{K}^{2} + \mathcal{O}(\beta^2)  \;, 
\end{equation}
and $a_{P}=m\ell_{P}^{2}/\hbar^{2}=(m/m_P)e_P^{-1}$, with $m_P$ and $e_P$ respectively the Planck mass and energy. 

The term $\hat{H}_{\beta}$ appearing in Eq.~\eqref{ModDyn} gives rise to a number of effects that can be observed experimentally also at low energies. For example, in the case of an oscillator, the term $\hat{H}_\beta\sim\op{p}^4$ results in an effective anharmonicity, and thus in an amplitude-dependent oscillation frequency. These effects have been probed experimentally on classical mechanical oscillators \cite{bawaj15,Tobar19,kumarNC20,marinNat13} and used to put upper bounds on $\beta$ under the assumption that a modified quantum algebra also affects the classical dynamics \cite{BenczikPRD02,Scardigli15,SCARDIGLI2017242,kumarNC20}.

While $\beta$ is typically considered to be a constant, it was recently proposed that, due to quantum gravity effects, this parameter could be stochastic and having the form of a white noise \cite{Petru21}. Here we generalize this model by including possible non-Markovian effects, \ie by assuming $\beta=\beta(t)$ being a stochastic variable characterized by
\begin{equation}\label{eq:BetaFl}
\avg{\beta(t)}=\overline{\beta} \;,\qquad \avg{(\beta(t)-\overline{\beta})(\beta(t')-\overline{\beta})}=\kappa f(t-t') \;, 
\end{equation}
where $\avg{...}$ denotes averaging over the stochastic process, $f(t-t')$ is a function satisfying $\int_{-\infty}^{+\infty} f(t)dt=1$ that characterizes the non-Markovianity of the model and $\kappa$ determines the amplitude of the fluctuations \footnote{While it is true that a time-dependent deformation parameter can break Lorentz invariance, it is important to remember that this is also the case for time-independent deformed commutators. Moreover, although attempts to derive Lorenz-invariant commutators exists \cite{Hossenfelder13}, it might be that this symmetry is anyways broken at the Planck scale \cite{maggiore93,SusskindPRD94}.}.
Different quantum gravity theories predict different values for $\beta$, but typically $\overline{\beta}\sim 1$. On the other hand, $\kappa$ was suggested to be the Planck time $t_P\approx 5.39\cdot\unit{10^{-44}}{s}$ \cite{Petru21}. However, since the exact value of these two parameters is unknown, they can be treated as free parameters that are bounded by experimental observations. Ultimately, the goal would be to put bounds so stringent that can rule out potential quantum gravity theories.

Equation~\eqref{ModDyn} with $\beta$ a random variable is now a stochastic Schr\"odinger equation, from which one can derive the corresponding
master equation (see SM \cite{SM}). Under the Born-Markov
approximation, one finds
\begin{align}
\partial_{t}\hat{\rho}(t) &= -\frac{i}{\hbar}\left[\hat{H}',\rho(t)\right] \notag\\
&- \frac{16a_{P}^{2}\kappa}{\hbar^{2}}\int_{0}^{t}dt'f(t-t')\left[\hat{K}^{2},\left[\hat{K}^{2I}(t'-t),\hat{\rho}(t)\right]\right] \;, \label{finalME}
\end{align}
where $\hat{H}'\equiv\hat{H}+4a_{P}\hat{K}^{2}\overline{\beta}$  

and $\hat{K}^{2I}(s)\equiv e^{i\hat{H}'s/\hbar}\hat{K}^{2}e^{-i\hat{H}'s/\hbar}$.
In deriving Eq.~(\ref{finalME}) we assumed that the correlation function $f(t)$ has a characteristic time $\tau$ such that $f(t\geq\tau)\simeq 0$ and that during the time $\tau$ the Hamiltonian evolution is negligible.

The first term in Eq.~\eqref{finalME} gives rise to a modified unitary evolution of the system, similar to the one mentioned for Eq.~\eqref{ModDyn}. The second term in Eq.~\eqref{finalME}, on the contrary, gives rise to a non-unitary evolution, \ie to a decoherence dynamics. Interestingly, this decoherence is in the momentum basis, which is also common in other proposal relating quantum mechanics and gravity \cite{karolyhazy1966gravitation,breuer2009metric, anastopoulos2013master,blencowe2013effective,bell2016quantum, bassi2017gravitational,gasbarri2017gravity,asprea2021gravitational,donadi2022seven}, although the dependence on $\hat{K}^2\sim \op{p}^4$ is a fingerprint of this model.

The Markovian limit of Eq.~\eqref{finalME} is obtained by setting $f(t-t')=\delta(t-t')$, which means that fluctuations of $\beta$ have an auto-correlation time small compared to other times scales. 
This simplifies $\hat{K}^{2I}(t'-t)\rightarrow \hat{K}^{2}$ and gives a factor $1/2$ for the time integral, leading to
\begin{equation}
\partial_{t}\hat{\rho}(t)=-\frac{i}{\hbar}\left[\hat{H}',\rho(t)\right]-\frac{8a_{P}^{2}\kappa}{\hbar^{2}}\left[\hat{K}^{2},\left[\hat{K}^{2},\hat{\rho}(t)\right]\right] \;. \label{finalMEmarkov} 
\end{equation}
This result agrees partially with the master equation derived in \cite{Petru21}, since there are three differences between our Eq.~\eqref{finalMEmarkov} and their Eq.~(19). 
The first one is the discrepancy of a factor 2 in the coefficient in front of the Lindblad term, which originates from the fact that in \cite{Petru21} it was not kept into account that the Dirac delta is centered at the extreme of the integration domain. 
The second one, more relevant, is that the unitary term is not just given by the system Hamiltonian $\hat{H}$, but it also contains extra terms proportional to $\overline{\beta}$ due to the modified commutator. These are expected to be there, since, in the limit of $\beta$ not fluctuating (i.e. $\kappa=0$), one still expects a correction to the unitary evolution consistent with Eq.~\eqref{ModDyn}. Finally, in Eq.~\eqref{finalMEmarkov} we have the parameter $\kappa$ in place of the Plank time present in Eq.~(19) of \cite{Petru21}, which we leave free to be bounded experimentally.

In the case of a free particle, $\hat{H}=\hat{K}=\hat{p}^2/2m$, Eq.~(\ref{finalME}) can be solved analytically in the momentum basis $\{\ket{p}\}$. By defining the matrix elements $\rho_{ab}(t):=\langle p_{a}|\hat{\rho}(t)|p_{b}\rangle$, we have
\begin{equation}\label{sol_free}
\rho_{ab}(t)=\rho_{ab}(0) \; e^{-\frac{i}{\hbar}\left(\Delta E_{1}+4a_{P}\Delta E_{2}\overline{\beta}\right)t}
\; e^{-(16a_{P}^{2}\kappa/\hbar^{2})\left( \Delta E_2 \right)^{2} g(t)} \;,
\end{equation}
with $\Delta E_k \equiv [(p^2_a/2m)^k-(p^2_b/2m)^k]$ and $g(t) \equiv \int_{0}^{t}dt'\int_{0}^{t'}dt''f(t'-t'')$.
The last term in Eq. (\ref{sol_free}) describes a damping of coherences between momentum eigenstates. 

\begin{figure}
    \centering
    \includegraphics[width=\columnwidth]{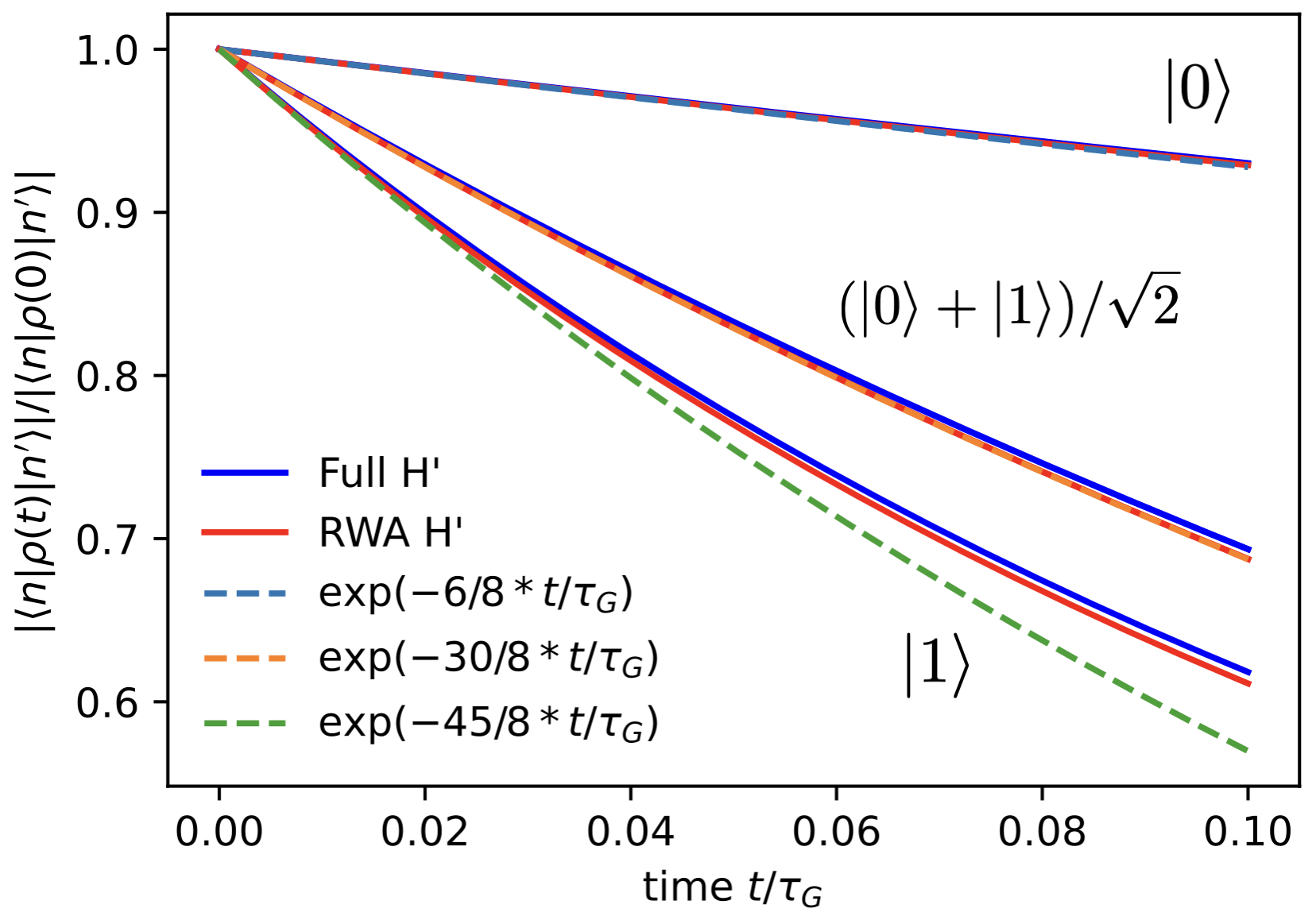}
    \caption{Numerical simulation of Eq.~\eqref{finalMEmarkov} (blue) and of Eq.~\eqref{HRWA} (red) for initial oscillator states $\ket{0}$, $(\ket{0}+\ket{1})/\sqrt{2}$ and $\ket{0}$, from top to bottom, compared to the analytical approximated results Eqs.~(\ref{r01final},\ref{overlaps}) (dashed). Note that for the two top-most cases all curves almost perfectly overlap. The simulation parameters $\overline{\beta}=1$, $\omega\, \tau_G=125\cdot 10^{3}$ and $\gamma=0$, are chosen to visualize the possible discrepancies between the different approximations.}
    \label{fig:sim}
\end{figure}

\vspace{2mm}
\textbf{Application to a harmonic oscillator.--}
We now want to investigate the dynamics resulting from Eq.~\eqref{finalME} when considering a harmonic oscillator, i.e. for $\hat{H}=\hat{K}+m\omega^2\op{x}^2/2$.
We start by applying the Rotating Wave Approximation (RWA) to $\hat{H}'=\hat{H}+4a_{P}\hat{K}^{2}\overline{\beta}$, which gives
\begin{equation}
\hat{H}'\simeq\hbar\omega\left(\hat{N}+\frac{1}{2}\right) + \dfrac{3}{8} a_{P}\overline{\beta}\hbar^{2}\omega^{2}\left(\hat{N}^{2}+\hat{N}+\frac{1}{2}\right)\equiv \hat{H}_{RWA}\label{HRWA}
\end{equation}
where $\hat{N}=\hat{a}^\dagger\hat{a}$ is the number operator with $\hat{a}$ and $\hat{a}^\dagger$ the usual (bosonic) ladder operators of the harmonic oscillator. 

At this point, let us remember that in a realistic experiment there are additional noise sources that contribute to the system decoherence. If these are well understood and characterized from independent measurements, they can be introduced as additional terms in Eq.~\eqref{finalME} to then obtain better bounds on $\kappa$. For example, of particular relevance for an oscillator is energy relaxation, which can be described by the Lindbladian $\gamma \hat{a}\hat{\rho}(t)\hat{a}^{\dagger}-\frac{\gamma}{2}\left\{ \hat{N},\hat{\rho}(t)\right\}$, where $\gamma$ is the energy relaxation rate. 
Taking this effect into account, Eq.~\eqref{finalME} becomes
\begin{align}\label{me_rwa_main2}
\partial_{t}\hat{\rho}(t)&=-\frac{i}{\hbar}\left[\hat{H}_{RWA},\hat{\rho}(t)\right] +\gamma \hat{a}\hat{\rho}(t)\hat{a}^{\dagger}-\frac{\gamma}{2}\left\{ \hat{N},\hat{\rho}(t)\right\} \\
&-\frac{16a_{P}^{2}\kappa}{\hbar^{2}}\int_{0}^{t}dt'f(t-t')\left[\hat{K}^{2},\left[\hat{K}^{I2}(t'-t),\hat{\rho}(t)\right]\right] \;. \nonumber
\end{align}
In the supplementary material, we solve Eq.~\eqref{me_rwa_main2} for different initial states of interest, by treating the term in the second line perturbatively. 
In what follows, we report the results for the Markovian case, for which the effects are stronger.

Taking as initial state $\hat{\rho}_{0+1}$ the superposition $(|0\rangle+|1\rangle)/\sqrt{2}$, we obtain that the coherence between Fock states $|0\rangle$ and $|1\rangle$ evolves as (see Eq. (\ref{ro01}) in the SM \cite{SM}) 
\begin{equation}\label{r01final}
\langle0|\hat{\rho}_{0+1}(t)|1\rangle \simeq \frac{1}{2} e^{-\frac{i}{\hbar}\left(E_{0}-E_{1}\right)t} e^{-\frac{\gamma}{2}t} \left(1-\dfrac{30}{8} \dfrac{t}{\tau_G} \right) \;,
\end{equation}
with $1/\tau_G \equiv 8 a_P^2 \kappa \hbar^2 \omega^4$. This result has been obtained by treating perturbatively the term in the last line of Eq. \eqref{me_rwa_main2}, meaning that it holds for $\omega\,  \tau_G,\gamma\,  \tau_G \gg 1$ and $t/\tau_G \ll 1$. 

Similarly to the free particle case Eq.~\eqref{sol_free}, we see from Eq.~\eqref{r01final} that a superposition of energy eigenstates experiences a decay due to the term in the second line of Eq.~\eqref{me_rwa_main2}. 
However, in the case of a harmonic oscillator, energy eigenstates are not eigenstates of the collapse operator $\op{K}^2\sim \op{p}^4$. For this reason, we observe a nontrivial evolution for any Fock state $\ket{n}$. For example (see Eqs. (\ref{ro11_sm}, \ref{ro00_sm}) in the SM \cite{SM})
\begin{equation}
    \langle 0|\hat{\rho}_{0}(t)|0\rangle \simeq \left(1-\dfrac{6}{8} \dfrac{t}{\tau_G} \right) ,\; \langle 1|\hat{\rho}_{1}(t)|1\rangle \simeq e^{-\gamma t} \left(1-\dfrac{45}{8} \dfrac{t}{\tau_G} \right)  \;, \label{overlaps}
\end{equation}
for initial states $\hat{\rho}_0=\ket{0}\bra{0}$ and $\hat{\rho}_1=\ket{1}\bra{1}$, respectively. Here, the decay noticeable in $\langle 0|\hat{\rho}_{0}(t)|0\rangle$ can be understood as a ``heating effect'' resulting from the collapse dynamics. For a comparison between Eqs.~(\ref{finalMEmarkov},\ref{HRWA}) and these analytical approximation for short-time evolution see Fig.~\ref{fig:sim}.

Importantly, through Eqs.~(\ref{r01final},\ref{overlaps}), an experimental measurement of $\bra{n}\hat{\rho}(t)\ket{n^\prime}$ can be used to put a bound on the fluctuation parameter $\kappa$. This is what we are going to show in the next section.

\vspace{2mm}
\textbf{Experimental tests.--} After having derived the decoherence dynamics resulting from the deformed commutator Eq.~\eqref{eq:ModComm} with fluctuating $\beta$, we now use experimental data from a massive oscillator in the quantum regime to put bounds on the parameters $\kappa$ and $\overline{\beta}$ of the model.

We consider measurements performed on a high-overtone bulk-acoustic wave resonator (HBAR) device, which makes use of a superconducting qubit to prepare, control, and read-out the quantum state of a vibration mode localized in the bulk of a sapphire crystal \cite{vonLupke22,catSCI23}. This mode is described to a very good approximation by a quantum harmonic oscillator with frequency $\omega = \unit{2\pi \cdot 5.96}{GHz}$ and effective mass $m_{\mathrm{eff}} = 16.2\,\mu$g \cite{catSCI23}, which corresponds to a zero-point fluctuation of $x_0=\sqrt{\hbar/2 m_{\mathrm{eff}}\omega}=\unit{2.9 \cdot 10^{-19}}{m}$ and to $a_P\hbar\omega=1.5\cdot 10^{-33}$. 

As the HBAR device is cooled down at a temperature of about $\unit{10}{mK}$, the high-frequency mechanical mode thermalizes very close to the ground state, with a residual steady-state population of $\avg{n} \leq 0.016(2)$ phonons. From this initial state, the interaction with the qubit allows us to prepare different nonclassical states of motion, such as Fock states \cite{vonLupke22}, Schr\"odinger cat states \cite{catSCI23}, and squeezed vacuum states \cite{squeezing23}. To investigate the predictions Eqs.~(\ref{r01final},\ref{overlaps}), here we are interested in the states $a\ket{0}+b\ket{1}$, which can be prepared by swapping the qubit state $a\ket{\downarrow}+b\ket{\uparrow}$ to the mechanical mode through the resonant Jaynes–Cummings interaction \cite{vonLupke22,macroPRL}.

\begin{figure*}[t!]
    \centering
    \includegraphics[width=0.9\textwidth]{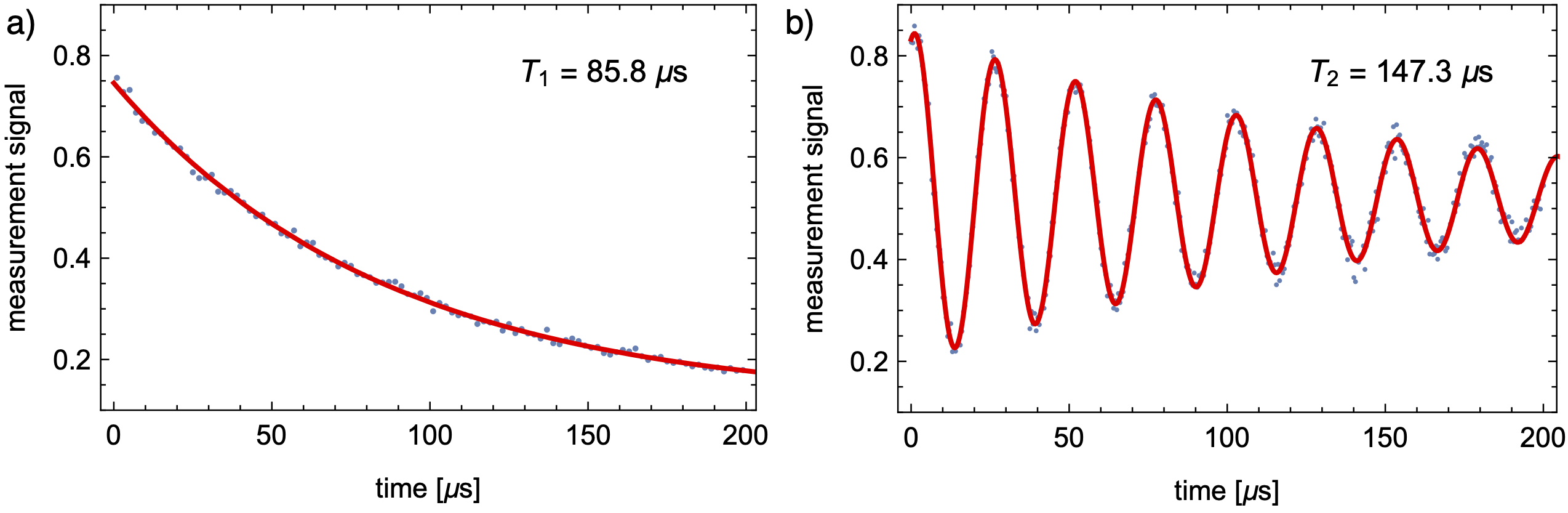}
    \caption{Left: measurement of the mechanical resonator energy relaxation rate. Fitting the data with a decaying exponential allows us to extract the $T_1$ time. Right: Ramsey measurement illustrating the decoherence of state $(\ket{0}+\ket{1})/\sqrt{2}$. Fitting the data with a decaying cosine allows us to extract the $T_2$ time. 
    See main text for details on the experimental sequence.}
    \label{fig1}
\end{figure*}

We begin with preparing the phonon mode in Fock state $\ket{1}$, and then measuring the remaining population after a waiting time $t$. By repeating this measurement for different $t$ we obtain the data shown in Fig.~\ref{fig1}a, that are fitted to a decaying exponential to obtain the decay time constant $T_1=\unit{85.8{\pm 1.5}}{\mu s}$, where the error is one standard deviation given by the fitting function. Let us remember that, according to Eq.~\eqref{overlaps}, this $T_1$ time includes both a contribution from energy relaxation and a contribution from the modified dynamics.

To distinguish between these two effects we perform a Ramsey-type measurement where we prepare the mechanical oscillator in state $(\ket{0}+\ket{1})/\sqrt{2}$, and monitor its phase evolution. 
This measurement of the phase is done by swapping the phonon state back to the qubit after an evolution time $t$, and then measuring the qubit state in the Pauli $\sigma_x$ basis. By repeating this sequence for different times $t$ we obtain the data shown in Fig.~\ref{fig1}b, that are fitted to an exponentially decaying oscillation giving us the Ramsey dephasing time $T_2=\unit{147.3{\pm 2.6}}{\mu s}$. 
According to Eq.~\eqref{r01final}, the reduction in contrast noticeable in Fig.~\ref{fig1}b is again due to a combination of energy relaxation and modified dynamics. However, since these two effects have different contributions than in the previous case, we can solve Eqs.~(\ref{r01final},\ref{overlaps}) to obtain the estimates $\gamma^{-1} = \unit{169.9\pm 47.5}{\mu s}$ and $\tau_G = \unit{975.2\pm 237.4}{\mu s}$ \footnote{Note that the measurements in Fig.~\ref{fig1} are taken for $t/\tau_G < 0.2$, which is compatible with the assumption $t/\tau_G \ll 1$ used to derive Eqs.~(\ref{r01final},\ref{overlaps}).}. 
Let us note that these numbers should be considered as upper bounds for the corresponding parameters: the dephasing time $T_2$ has certainly additional contributions from technical noise terms that have not been characterized independently. Therefore, attributing all the decay in Fig.~\ref{fig1}b to $\gamma$ and $\tau_G$ results in overestimating these parameters. 
Nevertheless, we can follow this conservative approach, and obtain the bound $\kappa \leq \unit{(4.0\pm 0.9) \cdot 10^{46}}{s}$. This number, which is much larger than the Planck time, highlights the difficulty of resolving fluctuations in the minimal length or, in other words, it indicates that such fluctuations are not likely to be the main responsible for the decoherence of massive quantum systems. For these to be resolved, $\kappa\sim t_P$, one would need $m^2 \omega^4 \tau_G \sim \unit{10^{113}}{Kg^2/s^3}$.

\vspace{2mm}
\textbf{Ground state deformation.--} 
When applied to the harmonic oscillator, the modified commutator Eq.~\eqref{eq:ModComm} with constant $\beta\neq 0$ results in a number of observable effects \cite{Scardigli_2019}. Among these there is a modification to the wavefunctions of the system \cite{SolHOscPRD,DadicPRD03}, which for the ground state takes the form of a ``squeezing'' originating from the $\op{p}^4$ nonlinearity appearing in the Hamiltonian Eq.~\eqref{Hb}.
Defining the measurement quadrature $\op{x}_{\theta} = \op{x} \cos\theta + \op{p} \sin\theta$, the ground state variance takes the form $ \Delta \op{x}_{\theta}^2 = \frac{1}{2} - \frac{1}{4} \epsilon \cos(2\theta)$ where $\epsilon \equiv 6 \overline{\beta} a_P \hbar\omega$.
Therefore, a precise measurement of the ground state ellipticity can be used to put a bound on $\overline{\beta}$. 

\begin{figure}[t!]
    \centering
    \includegraphics[width=0.78\columnwidth]{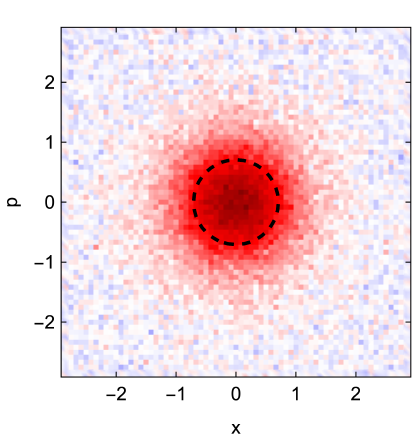}
    \caption{Wigner function measurement of the mechanical resonator ground state. Diagonalization of the associated covariance matrix allows us to estimate the state ellipticity. Dashed black circle represent one standard deviation of the ideal harmonic oscillator ground state.}
    \label{fig:2}
\end{figure}

We perform such a measurement in the considered mechanical oscillator by first taking the Wigner function of the ground state \cite{squeezing23}, Fig.~\ref{fig:2}, and then extracting the variances from a two-dimensional Gaussian fit. The ratio between the maximum (\ie $\theta=\pi/2$) and minimum (\ie $\theta=0$) variances is given by $\Delta\op{x}_{\text{max}}^2/\Delta\op{x}_{\text{min}}^2=(2+\epsilon)/(2-\epsilon)$, which allows us to extract $\epsilon=0.020(5)$. From this we obtain the bound $\overline{\beta}<2.2(6)\cdot 10^{30}$, which is tighter than the other also quantum bound $\overline{\beta}< 10^{34}$ obtained from the 1S-2S energy-level difference in hydrogen \cite{Quesne10}. Let us emphasize that, although our bound is less tight than other bounds obtained using mechanical oscillators \cite{bawaj15,Tobar19,kumarNC20}, this is the only one in which the oscillator operates in the quantum regime. Therefore, our result does not rely on the additional assumption that a deformed commutator result in deformed Poisson brackets.

As a side note, the same analysis of the ground state variances can be used to bound the nonlocality length scale $l_k$, namely the scale at which quantum gravity effects are expected to modify the standard local dynamics \cite{BelenchiaPRL}. In this framework, the effective (non-Hermitian) Hamiltonian takes the form $H+\epsilon(x^4/4+i\{x,p\})$ \cite{BelenchiaHam}, where $\epsilon\equiv l_k^2/x_0^2$. The quartic term $\sim \hat{x}^4$ plays a similar role of the term $\sim \hat{p}^4$ in $\hat{H}_\beta$ above, hence one can set a bound on $\epsilon$ and hence on $l_k$ by studying the ground state
variances. The measurements we consider results in the bound $l_k \leq 5.9(8)\cdot\unit{10^{-20}}{m}$. Remarkably, this is on par with the bound $l_k\leq 10^{-19}$ obtained in Ref.~\cite{Biswas15} from \unit{8}{TeV} LHC data. In fact, the dependence of $\epsilon$ on the mass of the oscillator, $\epsilon\sim m$, suggests that massive quantum systems are ideal platforms for putting stringent bounds on $l_k$.

\vspace{2mm}
\textbf{Metric fluctuations model.--} Our analysis can be applied to other models of decoherence due to spacetime fluctuations, such as the one suggested by Breuer \textit{et al.} in \cite{breuer2009metric}.
According to this model, the local metric tensor $g_{\mu\nu}(x,t)$ is the sum of a Minkowskian background $\eta_{\mu\nu}=\text{diag}(-1,1,1,1)$ and a small fluctuating part $|h_{\mu\nu}(x,t)|\ll 1$ that will be considered isotropic.
Taking a non–relativistic limit of the Klein–Gordon equation one obtains an effective Schr\"{o}dinger equation of the form Eq.~\eqref{ModDyn}, where the term $\hat{H}_\beta$ is now replaced by $\hat{H}_B = \sqrt{\tau_c} w(t) \hat{K}$, with $w(t)$ being a white noise \ie $\langle w(t)\rangle=0$ and $\langle w(t)w(t') \rangle=\delta(t-t')$. This results in the master equation \cite{breuer2009metric}
\begin{equation}
\frac{d\hat{\rho}(t)}{dt}=-\frac{i}{\hbar}\left[\hat{H},\hat{\rho}(t)\right]-\frac{\tau_c}{2\hbar^2}\left[\hat{K},\left[\hat{K},\hat{\rho}(t)\right]\right] \;. 
\end{equation}
To appreciate the difference between this result and Eq.~\eqref{finalMEmarkov}, it is instructive to observe that the decoherence of a free particle now reads 
\begin{equation}
\rho_{ab}(t)=\rho_{ab}(0)e^{-\frac{i}{\hbar} \Delta E_{1}t}e^{-(\tau_{c}/2\hbar^{2}) (\Delta E_{1})^{2}t} \;.
\end{equation}
Note that, contrary to Eq.~\eqref{sol_free}, the second exponential depends on $(\Delta E_{1})^{2}$, and not on $(\Delta E_{2})^{2}$.

Following the same steps we used to derive Eqs.~(\ref{r01final},\ref{overlaps}), we obtain for a harmonic oscillator (see SM \cite{SM})
\begin{equation}\label{AAAAAAAAAA}
\langle0|\hat{\rho}_{0+1}(t)|1\rangle \simeq \frac{1}{2} e^{-\frac{i}{\hbar}\left(E_{0}-E_{1}\right)t} e^{-\frac{\gamma}{2}t} \left(1-\dfrac{3}{8} \dfrac{t}{\tau_D} \right) \;,
\end{equation}
\begin{equation}
    \langle 0|\hat{\rho}_{0}(t)|0\rangle \simeq \left(1-\dfrac{1}{8} \dfrac{t}{\tau_D} \right) ,\; \langle 1|\hat{\rho}_{1}(t)|1\rangle \simeq e^{-\gamma t} \left(1-\dfrac{3}{8} \dfrac{t}{\tau_D} \right)  \;, \label{BBBBBBB}
\end{equation}
where $1/\tau_D \equiv \tau_c \omega^2$. Solving these two expressions with the data presented in Fig.~\ref{fig1} gives us $\gamma^{-1} = \unit{102.8\pm 6.9}{\mu s}$ and $\tau_D = \unit{195.0\pm 47.5}{\mu s}$, which correspond to the bound $\tau_c \leq \unit{(3.7\pm 0.8) \cdot 10^{-18}}{s}$. 
This bound is much tighter than the one on $\kappa$ because $\tau_D$ does not include the term $(a_P \hbar\omega)^{-2}$. Still, if we would like to obtain a bound $\tau_c \sim t_P$ it is necessary to have an oscillator with  $\gamma^{-1} \sim \tau_D$, meaning $\omega^2/\gamma \sim \unit{10^{43}}{s^{-1}}$ \footnote{Note that $\omega^2/\gamma \sim Q f $, the product of the oscillator's quality factor and frequency.}. 
However, note that there is no a priory reason to expect $\tau_c \sim t_P$, since the physical origin of metric fluctuations is not known and thus they might originate from processes at energy scales much lower than the Planck scale.

\vspace{2mm}
\textbf{Conclusions.--} We have investigated two models of fluctuating spacetimes, one originating from a deformed commutator with a fluctuating deformation parameter and the other originating from metric tensor fluctuations. We have derived the master equations describing the evolution of a quantum system in these scenarios, correcting mistakes in the literature, and discussed the resulting decoherence in momentum basis.
In particular, we computed how density matrix elements for free particles and harmonic oscillators decay.
Using experimental data taken from a \unit{16}{\mu g} mechanical oscillator prepared in quantum states of motion we bound the free parameters of the models, namely the fluctuations amplitude and mean value. We conclude that, while the effects of a fluctuating deformed commutator are extremely small to play an role at low energy scales, metric fluctuations have much stronger effects. 
In addition, the measurements we have presented can provide a bound to the commutator deformation parameter and to the nonlocality length scale which do not require additional assumptions since, contrary to other experiments with mechanical oscillators, our tests are in the quantum regime.

\vspace{2mm}
\textit{Acknowledgments.--} 
S.D. acknowledges support from the UKRI through grant EP/X021505/1. 
M.F. was supported by the Swiss National Science Foundation Ambizione Grant No. 208886, and by The Branco Weiss Fellowship -- Society in Science, administered by the ETH Z\"{u}rich. The authors would like to acknowledge the contribution of the COST Action CA23130 BridgeQG (Bridging High and Low Energies in Search for Quantum Gravity).

\bibliographystyle{apsrev4-1} 
\bibliography{mybib.bib}

\begin{thebibliography}{49}%
\makeatletter
\providecommand \@ifxundefined [1]{%
 \@ifx{#1\undefined}
}%
\providecommand \@ifnum [1]{%
 \ifnum #1\expandafter \@firstoftwo
 \else \expandafter \@secondoftwo
 \fi
}%
\providecommand \@ifx [1]{%
 \ifx #1\expandafter \@firstoftwo
 \else \expandafter \@secondoftwo
 \fi
}%
\providecommand \natexlab [1]{#1}%
\providecommand \enquote  [1]{``#1''}%
\providecommand \bibnamefont  [1]{#1}%
\providecommand \bibfnamefont [1]{#1}%
\providecommand \citenamefont [1]{#1}%
\providecommand \href@noop [0]{\@secondoftwo}%
\providecommand \href [0]{\begingroup \@sanitize@url \@href}%
\providecommand \@href[1]{\@@startlink{#1}\@@href}%
\providecommand \@@href[1]{\endgroup#1\@@endlink}%
\providecommand \@sanitize@url [0]{\catcode `\\12\catcode `\$12\catcode
  `\&12\catcode `\#12\catcode `\^12\catcode `\_12\catcode `\%12\relax}%
\providecommand \@@startlink[1]{}%
\providecommand \@@endlink[0]{}%
\providecommand \url  [0]{\begingroup\@sanitize@url \@url }%
\providecommand \@url [1]{\endgroup\@href {#1}{\urlprefix }}%
\providecommand \urlprefix  [0]{URL }%
\providecommand \Eprint [0]{\href }%
\providecommand \doibase [0]{http://dx.doi.org/}%
\providecommand \selectlanguage [0]{\@gobble}%
\providecommand \bibinfo  [0]{\@secondoftwo}%
\providecommand \bibfield  [0]{\@secondoftwo}%
\providecommand \translation [1]{[#1]}%
\providecommand \BibitemOpen [0]{}%
\providecommand \bibitemStop [0]{}%
\providecommand \bibitemNoStop [0]{.\EOS\space}%
\providecommand \EOS [0]{\spacefactor3000\relax}%
\providecommand \BibitemShut  [1]{\csname bibitem#1\endcsname}%
\let\auto@bib@innerbib\@empty
\bibitem [{\citenamefont {Veneziano}(1986)}]{veneziano_stringy_1986}%
  \BibitemOpen
  \bibfield  {author} {\bibinfo {author} {\bibfnamefont {G.}~\bibnamefont
  {Veneziano}},\ }\href {\doibase 10.1209/0295-5075/2/3/006} {\bibfield
  {journal} {\bibinfo  {journal} {Europhysics Letters ({EPL})}\ }\textbf
  {\bibinfo {volume} {2}},\ \bibinfo {pages} {199} (\bibinfo {year}
  {1986})}\BibitemShut {NoStop}%
\bibitem [{\citenamefont {Amati}\ \emph {et~al.}(1989)\citenamefont {Amati},
  \citenamefont {Ciafaloni},\ and\ \citenamefont {Veneziano}}]{amati_can_1989}%
  \BibitemOpen
  \bibfield  {author} {\bibinfo {author} {\bibfnamefont {D.}~\bibnamefont
  {Amati}}, \bibinfo {author} {\bibfnamefont {M.}~\bibnamefont {Ciafaloni}}, \
  and\ \bibinfo {author} {\bibfnamefont {G.}~\bibnamefont {Veneziano}},\ }\href
  {\doibase 10.1016/0370-2693(89)91366-X} {\bibfield  {journal} {\bibinfo
  {journal} {Physics Letters B}\ }\textbf {\bibinfo {volume} {216}},\ \bibinfo
  {pages} {41} (\bibinfo {year} {1989})}\BibitemShut {NoStop}%
\bibitem [{\citenamefont {Konishi}\ \emph {et~al.}(1990)\citenamefont
  {Konishi}, \citenamefont {Paffuti},\ and\ \citenamefont
  {Provero}}]{konishi_minimum_1990}%
  \BibitemOpen
  \bibfield  {author} {\bibinfo {author} {\bibfnamefont {K.}~\bibnamefont
  {Konishi}}, \bibinfo {author} {\bibfnamefont {G.}~\bibnamefont {Paffuti}}, \
  and\ \bibinfo {author} {\bibfnamefont {P.}~\bibnamefont {Provero}},\ }\href
  {\doibase 10.1016/0370-2693(90)91927-4} {\bibfield  {journal} {\bibinfo
  {journal} {Physics Letters B}\ }\textbf {\bibinfo {volume} {234}},\ \bibinfo
  {pages} {276} (\bibinfo {year} {1990})}\BibitemShut {NoStop}%
\bibitem [{\citenamefont
  {Maggiore}(1993{\natexlab{a}})}]{maggiore_generalized_1993}%
  \BibitemOpen
  \bibfield  {author} {\bibinfo {author} {\bibfnamefont {M.}~\bibnamefont
  {Maggiore}},\ }\href {\doibase 10.1016/0370-2693(93)91401-8} {\bibfield
  {journal} {\bibinfo  {journal} {Physics Letters B}\ }\textbf {\bibinfo
  {volume} {304}},\ \bibinfo {pages} {65} (\bibinfo {year}
  {1993}{\natexlab{a}})}\BibitemShut {NoStop}%
\bibitem [{\citenamefont {Garay}(1995)}]{garay_quantum_1995}%
  \BibitemOpen
  \bibfield  {author} {\bibinfo {author} {\bibfnamefont {L.~J.}\ \bibnamefont
  {Garay}},\ }\href {\doibase 10.1142/S0217751X95000085} {\bibfield  {journal}
  {\bibinfo  {journal} {International Journal of Modern Physics A}\ }\textbf
  {\bibinfo {volume} {10}},\ \bibinfo {pages} {145} (\bibinfo {year}
  {1995})}\BibitemShut {NoStop}%
\bibitem [{\citenamefont {Scardigli}(1999)}]{SCARDIGLI199939}%
  \BibitemOpen
  \bibfield  {author} {\bibinfo {author} {\bibfnamefont {F.}~\bibnamefont
  {Scardigli}},\ }\href {\doibase
  https://doi.org/10.1016/S0370-2693(99)00167-7} {\bibfield  {journal}
  {\bibinfo  {journal} {Physics Letters B}\ }\textbf {\bibinfo {volume}
  {452}},\ \bibinfo {pages} {39} (\bibinfo {year} {1999})}\BibitemShut
  {NoStop}%
\bibitem [{\citenamefont {Adler}\ and\ \citenamefont
  {Santiago}(1999)}]{Adler99}%
  \BibitemOpen
  \bibfield  {author} {\bibinfo {author} {\bibfnamefont {R.~J.}\ \bibnamefont
  {Adler}}\ and\ \bibinfo {author} {\bibfnamefont {D.~I.}\ \bibnamefont
  {Santiago}},\ }\href {\doibase 10.1142/S0217732399001462} {\bibfield
  {journal} {\bibinfo  {journal} {Modern Physics Letters A}\ }\textbf {\bibinfo
  {volume} {14}},\ \bibinfo {pages} {1371} (\bibinfo {year}
  {1999})}\BibitemShut {NoStop}%
\bibitem [{\citenamefont {Hossenfelder}(2013)}]{Hossenfelder13}%
  \BibitemOpen
  \bibfield  {author} {\bibinfo {author} {\bibfnamefont {S.}~\bibnamefont
  {Hossenfelder}},\ }\href {\doibase 10.12942/lrr-2013-2} {\bibfield  {journal}
  {\bibinfo  {journal} {Living Reviews in Relativity}\ }\textbf {\bibinfo
  {volume} {16}},\ \bibinfo {pages} {2} (\bibinfo {year} {2013})}\BibitemShut
  {NoStop}%
\bibitem [{\citenamefont {Maggiore}(1993{\natexlab{b}})}]{maggiore93}%
  \BibitemOpen
  \bibfield  {author} {\bibinfo {author} {\bibfnamefont {M.}~\bibnamefont
  {Maggiore}},\ }\href {\doibase https://doi.org/10.1016/0370-2693(93)90785-G}
  {\bibfield  {journal} {\bibinfo  {journal} {Physics Letters B}\ }\textbf
  {\bibinfo {volume} {319}},\ \bibinfo {pages} {83} (\bibinfo {year}
  {1993}{\natexlab{b}})}\BibitemShut {NoStop}%
\bibitem [{\citenamefont {Kempf}(1997)}]{Kempf97}%
  \BibitemOpen
  \bibfield  {author} {\bibinfo {author} {\bibfnamefont {A.}~\bibnamefont
  {Kempf}},\ }\href {\doibase 10.1088/0305-4470/30/6/030} {\bibfield  {journal}
  {\bibinfo  {journal} {Journal of Physics A: Mathematical and General}\
  }\textbf {\bibinfo {volume} {30}},\ \bibinfo {pages} {2093} (\bibinfo {year}
  {1997})}\BibitemShut {NoStop}%
\bibitem [{\citenamefont {Fadel}\ and\ \citenamefont
  {Maggiore}(2022)}]{fadelPRD22}%
  \BibitemOpen
  \bibfield  {author} {\bibinfo {author} {\bibfnamefont {M.}~\bibnamefont
  {Fadel}}\ and\ \bibinfo {author} {\bibfnamefont {M.}~\bibnamefont
  {Maggiore}},\ }\href {\doibase 10.1103/PhysRevD.105.106017} {\bibfield
  {journal} {\bibinfo  {journal} {Phys. Rev. D}\ }\textbf {\bibinfo {volume}
  {105}},\ \bibinfo {pages} {106017} (\bibinfo {year} {2022})}\BibitemShut
  {NoStop}%
\bibitem [{\citenamefont {Petruzziello}\ and\ \citenamefont
  {Illuminati}(2021)}]{Petru21}%
  \BibitemOpen
  \bibfield  {author} {\bibinfo {author} {\bibfnamefont {L.}~\bibnamefont
  {Petruzziello}}\ and\ \bibinfo {author} {\bibfnamefont {F.}~\bibnamefont
  {Illuminati}},\ }\href {\doibase 10.1038/s41467-021-24711-7} {\bibfield
  {journal} {\bibinfo  {journal} {Nature Communications}\ }\textbf {\bibinfo
  {volume} {12}},\ \bibinfo {pages} {4449} (\bibinfo {year}
  {2021})}\BibitemShut {NoStop}%
\bibitem [{\citenamefont {Das}\ \emph {et~al.}(2016)\citenamefont {Das},
  \citenamefont {Robbins},\ and\ \citenamefont {Walton}}]{Das16}%
  \BibitemOpen
  \bibfield  {author} {\bibinfo {author} {\bibfnamefont {S.}~\bibnamefont
  {Das}}, \bibinfo {author} {\bibfnamefont {M.~P.}\ \bibnamefont {Robbins}}, \
  and\ \bibinfo {author} {\bibfnamefont {M.~A.}\ \bibnamefont {Walton}},\
  }\href {\doibase 10.1139/cjp-2015-0456} {\bibfield  {journal} {\bibinfo
  {journal} {Canadian Journal of Physics}\ }\textbf {\bibinfo {volume} {94}},\
  \bibinfo {pages} {139} (\bibinfo {year} {2016})}\BibitemShut {NoStop}%
\bibitem [{\citenamefont {Ali}\ \emph {et~al.}(2011)\citenamefont {Ali},
  \citenamefont {Das},\ and\ \citenamefont {Vagenas}}]{AliPRD11}%
  \BibitemOpen
  \bibfield  {author} {\bibinfo {author} {\bibfnamefont {A.~F.}\ \bibnamefont
  {Ali}}, \bibinfo {author} {\bibfnamefont {S.}~\bibnamefont {Das}}, \ and\
  \bibinfo {author} {\bibfnamefont {E.~C.}\ \bibnamefont {Vagenas}},\ }\href
  {\doibase 10.1103/PhysRevD.84.044013} {\bibfield  {journal} {\bibinfo
  {journal} {Phys. Rev. D}\ }\textbf {\bibinfo {volume} {84}},\ \bibinfo
  {pages} {044013} (\bibinfo {year} {2011})}\BibitemShut {NoStop}%
\bibitem [{\citenamefont {Pedram}(2012)}]{Pedram12}%
  \BibitemOpen
  \bibfield  {author} {\bibinfo {author} {\bibfnamefont {P.}~\bibnamefont
  {Pedram}},\ }\href {\doibase 10.1103/PhysRevD.85.024016} {\bibfield
  {journal} {\bibinfo  {journal} {Phys. Rev. D}\ }\textbf {\bibinfo {volume}
  {85}},\ \bibinfo {pages} {024016} (\bibinfo {year} {2012})}\BibitemShut
  {NoStop}%
\bibitem [{\citenamefont {Bosso}\ \emph {et~al.}(2017)\citenamefont {Bosso},
  \citenamefont {Das},\ and\ \citenamefont {Mann}}]{Bosso17}%
  \BibitemOpen
  \bibfield  {author} {\bibinfo {author} {\bibfnamefont {P.}~\bibnamefont
  {Bosso}}, \bibinfo {author} {\bibfnamefont {S.}~\bibnamefont {Das}}, \ and\
  \bibinfo {author} {\bibfnamefont {R.~B.}\ \bibnamefont {Mann}},\ }\href
  {\doibase 10.1103/PhysRevD.96.066008} {\bibfield  {journal} {\bibinfo
  {journal} {Phys. Rev. D}\ }\textbf {\bibinfo {volume} {96}},\ \bibinfo
  {pages} {066008} (\bibinfo {year} {2017})}\BibitemShut {NoStop}%
\bibitem [{\citenamefont {Pikovski}\ \emph {et~al.}(2012)\citenamefont
  {Pikovski}, \citenamefont {Vanner}, \citenamefont {Aspelmeyer}, \citenamefont
  {Kim},\ and\ \citenamefont {Brukner}}]{IgorGUP12}%
  \BibitemOpen
  \bibfield  {author} {\bibinfo {author} {\bibfnamefont {I.}~\bibnamefont
  {Pikovski}}, \bibinfo {author} {\bibfnamefont {M.~R.}\ \bibnamefont
  {Vanner}}, \bibinfo {author} {\bibfnamefont {M.}~\bibnamefont {Aspelmeyer}},
  \bibinfo {author} {\bibfnamefont {M.~S.}\ \bibnamefont {Kim}}, \ and\
  \bibinfo {author} {\bibfnamefont {C.}~\bibnamefont {Brukner}},\ }\href
  {\doibase 10.1038/nphys2262} {\bibfield  {journal} {\bibinfo  {journal}
  {Nature Physics}\ }\textbf {\bibinfo {volume} {8}},\ \bibinfo {pages} {393}
  (\bibinfo {year} {2012})}\BibitemShut {NoStop}%
\bibitem [{\citenamefont {Bawaj}\ \emph {et~al.}(2015)\citenamefont {Bawaj},
  \citenamefont {Biancofiore}, \citenamefont {Bonaldi}, \citenamefont
  {Bonfigli}, \citenamefont {Borrielli}, \citenamefont {Di~Giuseppe},
  \citenamefont {Marconi}, \citenamefont {Marino}, \citenamefont {Natali},
  \citenamefont {Pontin}, \citenamefont {Prodi}, \citenamefont {Serra},
  \citenamefont {Vitali},\ and\ \citenamefont {Marin}}]{bawaj15}%
  \BibitemOpen
  \bibfield  {author} {\bibinfo {author} {\bibfnamefont {M.}~\bibnamefont
  {Bawaj}}, \bibinfo {author} {\bibfnamefont {C.}~\bibnamefont {Biancofiore}},
  \bibinfo {author} {\bibfnamefont {M.}~\bibnamefont {Bonaldi}}, \bibinfo
  {author} {\bibfnamefont {F.}~\bibnamefont {Bonfigli}}, \bibinfo {author}
  {\bibfnamefont {A.}~\bibnamefont {Borrielli}}, \bibinfo {author}
  {\bibfnamefont {G.}~\bibnamefont {Di~Giuseppe}}, \bibinfo {author}
  {\bibfnamefont {L.}~\bibnamefont {Marconi}}, \bibinfo {author} {\bibfnamefont
  {F.}~\bibnamefont {Marino}}, \bibinfo {author} {\bibfnamefont
  {R.}~\bibnamefont {Natali}}, \bibinfo {author} {\bibfnamefont
  {A.}~\bibnamefont {Pontin}}, \bibinfo {author} {\bibfnamefont {G.~A.}\
  \bibnamefont {Prodi}}, \bibinfo {author} {\bibfnamefont {E.}~\bibnamefont
  {Serra}}, \bibinfo {author} {\bibfnamefont {D.}~\bibnamefont {Vitali}}, \
  and\ \bibinfo {author} {\bibfnamefont {F.}~\bibnamefont {Marin}},\ }\href
  {\doibase 10.1038/ncomms8503} {\bibfield  {journal} {\bibinfo  {journal}
  {Nature Communications}\ }\textbf {\bibinfo {volume} {6}},\ \bibinfo {pages}
  {7503} (\bibinfo {year} {2015})}\BibitemShut {NoStop}%
\bibitem [{\citenamefont {Bushev}\ \emph {et~al.}(2019)\citenamefont {Bushev},
  \citenamefont {Bourhill}, \citenamefont {Goryachev}, \citenamefont
  {Kukharchyk}, \citenamefont {Ivanov}, \citenamefont {Galliou}, \citenamefont
  {Tobar},\ and\ \citenamefont {Danilishin}}]{Tobar19}%
  \BibitemOpen
  \bibfield  {author} {\bibinfo {author} {\bibfnamefont {P.~A.}\ \bibnamefont
  {Bushev}}, \bibinfo {author} {\bibfnamefont {J.}~\bibnamefont {Bourhill}},
  \bibinfo {author} {\bibfnamefont {M.}~\bibnamefont {Goryachev}}, \bibinfo
  {author} {\bibfnamefont {N.}~\bibnamefont {Kukharchyk}}, \bibinfo {author}
  {\bibfnamefont {E.}~\bibnamefont {Ivanov}}, \bibinfo {author} {\bibfnamefont
  {S.}~\bibnamefont {Galliou}}, \bibinfo {author} {\bibfnamefont {M.~E.}\
  \bibnamefont {Tobar}}, \ and\ \bibinfo {author} {\bibfnamefont
  {S.}~\bibnamefont {Danilishin}},\ }\href {\doibase
  10.1103/PhysRevD.100.066020} {\bibfield  {journal} {\bibinfo  {journal}
  {Physical Review D}\ }\textbf {\bibinfo {volume} {100}},\ \bibinfo {pages}
  {066020} (\bibinfo {year} {2019})}\BibitemShut {NoStop}%
\bibitem [{\citenamefont {Marin}\ \emph {et~al.}(2013)\citenamefont {Marin},
  \citenamefont {Marino}, \citenamefont {Bonaldi}, \citenamefont {Cerdonio},
  \citenamefont {Conti}, \citenamefont {Falferi}, \citenamefont {Mezzena},
  \citenamefont {Ortolan}, \citenamefont {Prodi}, \citenamefont {Taffarello},
  \citenamefont {Vedovato}, \citenamefont {Vinante},\ and\ \citenamefont
  {Zendri}}]{marinNat13}%
  \BibitemOpen
  \bibfield  {author} {\bibinfo {author} {\bibfnamefont {F.}~\bibnamefont
  {Marin}}, \bibinfo {author} {\bibfnamefont {F.}~\bibnamefont {Marino}},
  \bibinfo {author} {\bibfnamefont {M.}~\bibnamefont {Bonaldi}}, \bibinfo
  {author} {\bibfnamefont {M.}~\bibnamefont {Cerdonio}}, \bibinfo {author}
  {\bibfnamefont {L.}~\bibnamefont {Conti}}, \bibinfo {author} {\bibfnamefont
  {P.}~\bibnamefont {Falferi}}, \bibinfo {author} {\bibfnamefont
  {R.}~\bibnamefont {Mezzena}}, \bibinfo {author} {\bibfnamefont
  {A.}~\bibnamefont {Ortolan}}, \bibinfo {author} {\bibfnamefont {G.~A.}\
  \bibnamefont {Prodi}}, \bibinfo {author} {\bibfnamefont {L.}~\bibnamefont
  {Taffarello}}, \bibinfo {author} {\bibfnamefont {G.}~\bibnamefont
  {Vedovato}}, \bibinfo {author} {\bibfnamefont {A.}~\bibnamefont {Vinante}}, \
  and\ \bibinfo {author} {\bibfnamefont {J.-P.}\ \bibnamefont {Zendri}},\
  }\href {\doibase 10.1038/nphys2503} {\bibfield  {journal} {\bibinfo
  {journal} {Nature Physics}\ }\textbf {\bibinfo {volume} {9}},\ \bibinfo
  {pages} {71} (\bibinfo {year} {2013})}\BibitemShut {NoStop}%
\bibitem [{\citenamefont {Benczik}\ \emph {et~al.}(2002)\citenamefont
  {Benczik}, \citenamefont {Chang}, \citenamefont {Minic}, \citenamefont
  {Okamura}, \citenamefont {Rayyan},\ and\ \citenamefont
  {Takeuchi}}]{BenczikPRD02}%
  \BibitemOpen
  \bibfield  {author} {\bibinfo {author} {\bibfnamefont {S.}~\bibnamefont
  {Benczik}}, \bibinfo {author} {\bibfnamefont {L.~N.}\ \bibnamefont {Chang}},
  \bibinfo {author} {\bibfnamefont {D.}~\bibnamefont {Minic}}, \bibinfo
  {author} {\bibfnamefont {N.}~\bibnamefont {Okamura}}, \bibinfo {author}
  {\bibfnamefont {S.}~\bibnamefont {Rayyan}}, \ and\ \bibinfo {author}
  {\bibfnamefont {T.}~\bibnamefont {Takeuchi}},\ }\href {\doibase
  10.1103/PhysRevD.66.026003} {\bibfield  {journal} {\bibinfo  {journal} {Phys.
  Rev. D}\ }\textbf {\bibinfo {volume} {66}},\ \bibinfo {pages} {026003}
  (\bibinfo {year} {2002})}\BibitemShut {NoStop}%
\bibitem [{\citenamefont {Scardigli}\ and\ \citenamefont
  {Casadio}(2015)}]{Scardigli15}%
  \BibitemOpen
  \bibfield  {author} {\bibinfo {author} {\bibfnamefont {F.}~\bibnamefont
  {Scardigli}}\ and\ \bibinfo {author} {\bibfnamefont {R.}~\bibnamefont
  {Casadio}},\ }\href {\doibase 10.1140/epjc/s10052-015-3635-y} {\bibfield
  {journal} {\bibinfo  {journal} {The European Physical Journal C}\ }\textbf
  {\bibinfo {volume} {75}},\ \bibinfo {pages} {425} (\bibinfo {year}
  {2015})}\BibitemShut {NoStop}%
\bibitem [{\citenamefont {Scardigli}\ \emph {et~al.}(2017)\citenamefont
  {Scardigli}, \citenamefont {Lambiase},\ and\ \citenamefont
  {Vagenas}}]{SCARDIGLI2017242}%
  \BibitemOpen
  \bibfield  {author} {\bibinfo {author} {\bibfnamefont {F.}~\bibnamefont
  {Scardigli}}, \bibinfo {author} {\bibfnamefont {G.}~\bibnamefont {Lambiase}},
  \ and\ \bibinfo {author} {\bibfnamefont {E.~C.}\ \bibnamefont {Vagenas}},\
  }\href {\doibase https://doi.org/10.1016/j.physletb.2017.01.054} {\bibfield
  {journal} {\bibinfo  {journal} {Physics Letters B}\ }\textbf {\bibinfo
  {volume} {767}},\ \bibinfo {pages} {242} (\bibinfo {year}
  {2017})}\BibitemShut {NoStop}%
\bibitem [{\citenamefont {Kumar}\ and\ \citenamefont
  {Plenio}(2020)}]{kumarNC20}%
  \BibitemOpen
  \bibfield  {author} {\bibinfo {author} {\bibfnamefont {S.~P.}\ \bibnamefont
  {Kumar}}\ and\ \bibinfo {author} {\bibfnamefont {M.~B.}\ \bibnamefont
  {Plenio}},\ }\href {\doibase 10.1038/s41467-020-17518-5} {\bibfield
  {journal} {\bibinfo  {journal} {Nature Communications}\ }\textbf {\bibinfo
  {volume} {11}},\ \bibinfo {pages} {3900} (\bibinfo {year}
  {2020})}\BibitemShut {NoStop}%
\bibitem [{\citenamefont {Belenchia}\ \emph {et~al.}(2016)\citenamefont
  {Belenchia}, \citenamefont {Benincasa}, \citenamefont {Liberati},
  \citenamefont {Marin}, \citenamefont {Marino},\ and\ \citenamefont
  {Ortolan}}]{BelenchiaPRL}%
  \BibitemOpen
  \bibfield  {author} {\bibinfo {author} {\bibfnamefont {A.}~\bibnamefont
  {Belenchia}}, \bibinfo {author} {\bibfnamefont {D.~M.~T.}\ \bibnamefont
  {Benincasa}}, \bibinfo {author} {\bibfnamefont {S.}~\bibnamefont {Liberati}},
  \bibinfo {author} {\bibfnamefont {F.}~\bibnamefont {Marin}}, \bibinfo
  {author} {\bibfnamefont {F.}~\bibnamefont {Marino}}, \ and\ \bibinfo {author}
  {\bibfnamefont {A.}~\bibnamefont {Ortolan}},\ }\href {\doibase
  10.1103/PhysRevLett.116.161303} {\bibfield  {journal} {\bibinfo  {journal}
  {Phys. Rev. Lett.}\ }\textbf {\bibinfo {volume} {116}},\ \bibinfo {pages}
  {161303} (\bibinfo {year} {2016})}\BibitemShut {NoStop}%
\bibitem [{\citenamefont {Breuer}\ \emph {et~al.}(2009)\citenamefont {Breuer},
  \citenamefont {Göklü},\ and\ \citenamefont
  {Lämmerzahl}}]{breuer2009metric}%
  \BibitemOpen
  \bibfield  {author} {\bibinfo {author} {\bibfnamefont {H.-P.}\ \bibnamefont
  {Breuer}}, \bibinfo {author} {\bibfnamefont {E.}~\bibnamefont {Göklü}}, \
  and\ \bibinfo {author} {\bibfnamefont {C.}~\bibnamefont {Lämmerzahl}},\
  }\href {\doibase 10.1088/0264-9381/26/10/105012} {\bibfield  {journal}
  {\bibinfo  {journal} {Classical and Quantum Gravity}\ }\textbf {\bibinfo
  {volume} {26}},\ \bibinfo {pages} {105012} (\bibinfo {year}
  {2009})}\BibitemShut {NoStop}%
\bibitem [{Note1()}]{Note1}%
  \BibitemOpen
  \bibinfo {note} {While it is true that a time-dependent deformation parameter
  can break Lorentz invariance, it is important to remember that this is also
  the case for time-independent deformed commutators. Moreover, although
  attempts to derive Lorenz-invariant commutators exists \cite
  {Hossenfelder13}, it might be that this symmetry is anyways broken at the
  Planck scale \cite {maggiore93,SusskindPRD94}.}\BibitemShut {Stop}%
\bibitem [{SM()}]{SM}%
  \BibitemOpen
  \href@noop {} {\bibinfo  {journal} {See supplementary materials}\
  }\BibitemShut {NoStop}%
\bibitem [{\citenamefont {Karolyhazy}(1966)}]{karolyhazy1966gravitation}%
  \BibitemOpen
\bibfield  {journal} {  }\bibfield  {author} {\bibinfo {author} {\bibfnamefont
  {F.}~\bibnamefont {Karolyhazy}},\ }\href {\doibase 10.1007/BF02717926}
  {\bibfield  {journal} {\bibinfo  {journal} {Il Nuovo Cimento A}\ }\textbf
  {\bibinfo {volume} {42}},\ \bibinfo {pages} {390} (\bibinfo {year}
  {1966})}\BibitemShut {NoStop}%
\bibitem [{\citenamefont {Anastopoulos}\ and\ \citenamefont
  {Hu}(2013)}]{anastopoulos2013master}%
  \BibitemOpen
  \bibfield  {author} {\bibinfo {author} {\bibfnamefont {C.}~\bibnamefont
  {Anastopoulos}}\ and\ \bibinfo {author} {\bibfnamefont {B.~L.}\ \bibnamefont
  {Hu}},\ }\href {\doibase 10.1088/0264-9381/30/16/165007} {\bibfield
  {journal} {\bibinfo  {journal} {Classical and Quantum Gravity}\ }\textbf
  {\bibinfo {volume} {30}},\ \bibinfo {pages} {165007} (\bibinfo {year}
  {2013})}\BibitemShut {NoStop}%
\bibitem [{\citenamefont {Blencowe}(2013)}]{blencowe2013effective}%
  \BibitemOpen
  \bibfield  {author} {\bibinfo {author} {\bibfnamefont {M.~P.}\ \bibnamefont
  {Blencowe}},\ }\href {\doibase 10.1103/PhysRevLett.111.021302} {\bibfield
  {journal} {\bibinfo  {journal} {Phys. Rev. Lett.}\ }\textbf {\bibinfo
  {volume} {111}},\ \bibinfo {pages} {021302} (\bibinfo {year}
  {2013})}\BibitemShut {NoStop}%
\bibitem [{\citenamefont {Adler}(2016)}]{bell2016quantum}%
  \BibitemOpen
  \bibfield  {author} {\bibinfo {author} {\bibfnamefont {S.}~\bibnamefont
  {Adler}},\ }\href@noop {} {\emph {\bibinfo {title} {Gravitation and the noise
  needed in objective reduction models}}}\ (\bibinfo  {publisher} {in Quantum
  Nonlocality and Reality: 50 Years of Bell's Theorem, ed Mary Bell and Shan
  Gao, Cambridge University Press},\ \bibinfo {year} {2016})\BibitemShut
  {NoStop}%
\bibitem [{\citenamefont {Bassi}\ \emph {et~al.}(2017)\citenamefont {Bassi},
  \citenamefont {Gro{\ss}ardt},\ and\ \citenamefont
  {Ulbricht}}]{bassi2017gravitational}%
  \BibitemOpen
  \bibfield  {author} {\bibinfo {author} {\bibfnamefont {A.}~\bibnamefont
  {Bassi}}, \bibinfo {author} {\bibfnamefont {A.}~\bibnamefont {Gro{\ss}ardt}},
  \ and\ \bibinfo {author} {\bibfnamefont {H.}~\bibnamefont {Ulbricht}},\
  }\href {\doibase 10.1088/1361-6382/aa864f} {\bibfield  {journal} {\bibinfo
  {journal} {Classical and Quantum Gravity}\ }\textbf {\bibinfo {volume}
  {34}},\ \bibinfo {pages} {193002} (\bibinfo {year} {2017})}\BibitemShut
  {NoStop}%
\bibitem [{\citenamefont {Gasbarri}\ \emph {et~al.}(2017)\citenamefont
  {Gasbarri}, \citenamefont {Toro\ifmmode~\check{s}\else \v{s}\fi{}},
  \citenamefont {Donadi},\ and\ \citenamefont {Bassi}}]{gasbarri2017gravity}%
  \BibitemOpen
  \bibfield  {author} {\bibinfo {author} {\bibfnamefont {G.}~\bibnamefont
  {Gasbarri}}, \bibinfo {author} {\bibfnamefont {M.}~\bibnamefont
  {Toro\ifmmode~\check{s}\else \v{s}\fi{}}}, \bibinfo {author} {\bibfnamefont
  {S.}~\bibnamefont {Donadi}}, \ and\ \bibinfo {author} {\bibfnamefont
  {A.}~\bibnamefont {Bassi}},\ }\href {\doibase 10.1103/PhysRevD.96.104013}
  {\bibfield  {journal} {\bibinfo  {journal} {Phys. Rev. D}\ }\textbf {\bibinfo
  {volume} {96}},\ \bibinfo {pages} {104013} (\bibinfo {year}
  {2017})}\BibitemShut {NoStop}%
\bibitem [{\citenamefont {Asprea}\ \emph {et~al.}(2021)\citenamefont {Asprea},
  \citenamefont {Gasbarri},\ and\ \citenamefont
  {Bassi}}]{asprea2021gravitational}%
  \BibitemOpen
  \bibfield  {author} {\bibinfo {author} {\bibfnamefont {L.}~\bibnamefont
  {Asprea}}, \bibinfo {author} {\bibfnamefont {G.}~\bibnamefont {Gasbarri}}, \
  and\ \bibinfo {author} {\bibfnamefont {A.}~\bibnamefont {Bassi}},\ }\href
  {\doibase 10.1103/PhysRevD.103.104041} {\bibfield  {journal} {\bibinfo
  {journal} {Physical Review D}\ }\textbf {\bibinfo {volume} {103}},\ \bibinfo
  {pages} {104041} (\bibinfo {year} {2021})}\BibitemShut {NoStop}%
\bibitem [{\citenamefont {Donadi}\ and\ \citenamefont
  {Bassi}(2022)}]{donadi2022seven}%
  \BibitemOpen
  \bibfield  {author} {\bibinfo {author} {\bibfnamefont {S.}~\bibnamefont
  {Donadi}}\ and\ \bibinfo {author} {\bibfnamefont {A.}~\bibnamefont {Bassi}},\
  }\href {\doibase 10.1116/5.0089318} {\bibfield  {journal} {\bibinfo
  {journal} {AVS Quantum Science}\ }\textbf {\bibinfo {volume} {4}},\ \bibinfo
  {pages} {025601} (\bibinfo {year} {2022})}\BibitemShut {NoStop}%
\bibitem [{\citenamefont {von L{\"u}pke}\ \emph {et~al.}(2022)\citenamefont
  {von L{\"u}pke}, \citenamefont {Yang}, \citenamefont {Bild}, \citenamefont
  {Michaud}, \citenamefont {Fadel},\ and\ \citenamefont {Chu}}]{vonLupke22}%
  \BibitemOpen
  \bibfield  {author} {\bibinfo {author} {\bibfnamefont {U.}~\bibnamefont {von
  L{\"u}pke}}, \bibinfo {author} {\bibfnamefont {Y.}~\bibnamefont {Yang}},
  \bibinfo {author} {\bibfnamefont {M.}~\bibnamefont {Bild}}, \bibinfo {author}
  {\bibfnamefont {L.}~\bibnamefont {Michaud}}, \bibinfo {author} {\bibfnamefont
  {M.}~\bibnamefont {Fadel}}, \ and\ \bibinfo {author} {\bibfnamefont
  {Y.}~\bibnamefont {Chu}},\ }\href {\doibase 10.1038/s41567-022-01591-2}
  {\bibfield  {journal} {\bibinfo  {journal} {Nature Physics}\ }\textbf
  {\bibinfo {volume} {18}},\ \bibinfo {pages} {794} (\bibinfo {year}
  {2022})}\BibitemShut {NoStop}%
\bibitem [{\citenamefont {Bild}\ \emph {et~al.}(2023)\citenamefont {Bild},
  \citenamefont {Fadel}, \citenamefont {Yang}, \citenamefont {von Lüpke},
  \citenamefont {Martin}, \citenamefont {Bruno},\ and\ \citenamefont
  {Chu}}]{catSCI23}%
  \BibitemOpen
  \bibfield  {author} {\bibinfo {author} {\bibfnamefont {M.}~\bibnamefont
  {Bild}}, \bibinfo {author} {\bibfnamefont {M.}~\bibnamefont {Fadel}},
  \bibinfo {author} {\bibfnamefont {Y.}~\bibnamefont {Yang}}, \bibinfo {author}
  {\bibfnamefont {U.}~\bibnamefont {von Lüpke}}, \bibinfo {author}
  {\bibfnamefont {P.}~\bibnamefont {Martin}}, \bibinfo {author} {\bibfnamefont
  {A.}~\bibnamefont {Bruno}}, \ and\ \bibinfo {author} {\bibfnamefont
  {Y.}~\bibnamefont {Chu}},\ }\href {\doibase 10.1126/science.adf7553}
  {\bibfield  {journal} {\bibinfo  {journal} {Science}\ }\textbf {\bibinfo
  {volume} {380}},\ \bibinfo {pages} {274} (\bibinfo {year}
  {2023})}\BibitemShut {NoStop}%
\bibitem [{\citenamefont {Marti}\ \emph {et~al.}(2024)\citenamefont {Marti},
  \citenamefont {Von~Lüpke}, \citenamefont {Joshi}, \citenamefont {Yang},
  \citenamefont {Bild}, \citenamefont {Omahen}, \citenamefont {Chu},\ and\
  \citenamefont {Fadel}}]{squeezing23}%
  \BibitemOpen
  \bibfield  {author} {\bibinfo {author} {\bibfnamefont {S.}~\bibnamefont
  {Marti}}, \bibinfo {author} {\bibfnamefont {U.}~\bibnamefont {Von~Lüpke}},
  \bibinfo {author} {\bibfnamefont {O.}~\bibnamefont {Joshi}}, \bibinfo
  {author} {\bibfnamefont {Y.}~\bibnamefont {Yang}}, \bibinfo {author}
  {\bibfnamefont {M.}~\bibnamefont {Bild}}, \bibinfo {author} {\bibfnamefont
  {A.}~\bibnamefont {Omahen}}, \bibinfo {author} {\bibfnamefont
  {Y.}~\bibnamefont {Chu}}, \ and\ \bibinfo {author} {\bibfnamefont
  {M.}~\bibnamefont {Fadel}},\ }\href {\doibase 10.1038/s41567-024-02545-6}
  {\bibfield  {journal} {\bibinfo  {journal} {Nature Physics}\ }\textbf
  {\bibinfo {volume} {20}},\ \bibinfo {pages} {1448} (\bibinfo {year}
  {2024})}\BibitemShut {NoStop}%
\bibitem [{\citenamefont {Schrinski}\ \emph {et~al.}(2023)\citenamefont
  {Schrinski}, \citenamefont {Yang}, \citenamefont {von L\"upke}, \citenamefont
  {Bild}, \citenamefont {Chu}, \citenamefont {Hornberger}, \citenamefont
  {Nimmrichter},\ and\ \citenamefont {Fadel}}]{macroPRL}%
  \BibitemOpen
  \bibfield  {author} {\bibinfo {author} {\bibfnamefont {B.}~\bibnamefont
  {Schrinski}}, \bibinfo {author} {\bibfnamefont {Y.}~\bibnamefont {Yang}},
  \bibinfo {author} {\bibfnamefont {U.}~\bibnamefont {von L\"upke}}, \bibinfo
  {author} {\bibfnamefont {M.}~\bibnamefont {Bild}}, \bibinfo {author}
  {\bibfnamefont {Y.}~\bibnamefont {Chu}}, \bibinfo {author} {\bibfnamefont
  {K.}~\bibnamefont {Hornberger}}, \bibinfo {author} {\bibfnamefont
  {S.}~\bibnamefont {Nimmrichter}}, \ and\ \bibinfo {author} {\bibfnamefont
  {M.}~\bibnamefont {Fadel}},\ }\href {\doibase 10.1103/PhysRevLett.130.133604}
  {\bibfield  {journal} {\bibinfo  {journal} {Phys. Rev. Lett.}\ }\textbf
  {\bibinfo {volume} {130}},\ \bibinfo {pages} {133604} (\bibinfo {year}
  {2023})}\BibitemShut {NoStop}%
\bibitem [{Note2()}]{Note2}%
  \BibitemOpen
  \bibinfo {note} {Note that the measurements in Fig.~\ref {fig1} are taken for
  $t/\tau _G < 0.2$, which is compatible with the assumption $t/\tau _G \ll 1$
  used to derive Eqs.~(\ref {r01final},\ref {overlaps}).}\BibitemShut {Stop}%
\bibitem [{\citenamefont {Scardigli}(2019)}]{Scardigli_2019}%
  \BibitemOpen
  \bibfield  {author} {\bibinfo {author} {\bibfnamefont {F.}~\bibnamefont
  {Scardigli}},\ }\href {\doibase 10.1088/1742-6596/1275/1/012004} {\bibfield
  {journal} {\bibinfo  {journal} {Journal of Physics: Conference Series}\
  }\textbf {\bibinfo {volume} {1275}},\ \bibinfo {pages} {012004} (\bibinfo
  {year} {2019})}\BibitemShut {NoStop}%
\bibitem [{\citenamefont {Chang}\ \emph {et~al.}(2002)\citenamefont {Chang},
  \citenamefont {Minic}, \citenamefont {Okamura},\ and\ \citenamefont
  {Takeuchi}}]{SolHOscPRD}%
  \BibitemOpen
  \bibfield  {author} {\bibinfo {author} {\bibfnamefont {L.~N.}\ \bibnamefont
  {Chang}}, \bibinfo {author} {\bibfnamefont {D.}~\bibnamefont {Minic}},
  \bibinfo {author} {\bibfnamefont {N.}~\bibnamefont {Okamura}}, \ and\
  \bibinfo {author} {\bibfnamefont {T.}~\bibnamefont {Takeuchi}},\ }\href
  {\doibase 10.1103/PhysRevD.65.125027} {\bibfield  {journal} {\bibinfo
  {journal} {Phys. Rev. D}\ }\textbf {\bibinfo {volume} {65}},\ \bibinfo
  {pages} {125027} (\bibinfo {year} {2002})}\BibitemShut {NoStop}%
\bibitem [{\citenamefont {Dadi\ifmmode~\acute{c}\else \'{c}\fi{}}\ \emph
  {et~al.}(2003)\citenamefont {Dadi\ifmmode~\acute{c}\else \'{c}\fi{}},
  \citenamefont {Jonke},\ and\ \citenamefont {Meljanac}}]{DadicPRD03}%
  \BibitemOpen
  \bibfield  {author} {\bibinfo {author} {\bibfnamefont {I.}~\bibnamefont
  {Dadi\ifmmode~\acute{c}\else \'{c}\fi{}}}, \bibinfo {author} {\bibfnamefont
  {L.}~\bibnamefont {Jonke}}, \ and\ \bibinfo {author} {\bibfnamefont
  {S.}~\bibnamefont {Meljanac}},\ }\href {\doibase 10.1103/PhysRevD.67.087701}
  {\bibfield  {journal} {\bibinfo  {journal} {Phys. Rev. D}\ }\textbf {\bibinfo
  {volume} {67}},\ \bibinfo {pages} {087701} (\bibinfo {year}
  {2003})}\BibitemShut {NoStop}%
\bibitem [{\citenamefont {Quesne}\ and\ \citenamefont
  {Tkachuk}(2010)}]{Quesne10}%
  \BibitemOpen
  \bibfield  {author} {\bibinfo {author} {\bibfnamefont {C.}~\bibnamefont
  {Quesne}}\ and\ \bibinfo {author} {\bibfnamefont {V.~M.}\ \bibnamefont
  {Tkachuk}},\ }\href {\doibase 10.1103/PhysRevA.81.012106} {\bibfield
  {journal} {\bibinfo  {journal} {Phys. Rev. A}\ }\textbf {\bibinfo {volume}
  {81}},\ \bibinfo {pages} {012106} (\bibinfo {year} {2010})}\BibitemShut
  {NoStop}%
\bibitem [{\citenamefont {Belenchia}\ \emph {et~al.}(2019)\citenamefont
  {Belenchia}, \citenamefont {Benincasa}, \citenamefont {Marin}, \citenamefont
  {Marino}, \citenamefont {Ortolan}, \citenamefont {Paternostro},\ and\
  \citenamefont {Liberati}}]{BelenchiaHam}%
  \BibitemOpen
  \bibfield  {author} {\bibinfo {author} {\bibfnamefont {A.}~\bibnamefont
  {Belenchia}}, \bibinfo {author} {\bibfnamefont {D.}~\bibnamefont
  {Benincasa}}, \bibinfo {author} {\bibfnamefont {F.}~\bibnamefont {Marin}},
  \bibinfo {author} {\bibfnamefont {F.}~\bibnamefont {Marino}}, \bibinfo
  {author} {\bibfnamefont {A.}~\bibnamefont {Ortolan}}, \bibinfo {author}
  {\bibfnamefont {M.}~\bibnamefont {Paternostro}}, \ and\ \bibinfo {author}
  {\bibfnamefont {S.}~\bibnamefont {Liberati}},\ }\href {\doibase
  10.1088/1361-6382/ab2c0a} {\bibfield  {journal} {\bibinfo  {journal}
  {Classical and Quantum Gravity}\ }\textbf {\bibinfo {volume} {36}},\ \bibinfo
  {pages} {155006} (\bibinfo {year} {2019})}\BibitemShut {NoStop}%
\bibitem [{\citenamefont {Biswas}\ and\ \citenamefont
  {Okada}(2015)}]{Biswas15}%
  \BibitemOpen
  \bibfield  {author} {\bibinfo {author} {\bibfnamefont {T.}~\bibnamefont
  {Biswas}}\ and\ \bibinfo {author} {\bibfnamefont {N.}~\bibnamefont {Okada}},\
  }\href {\doibase https://doi.org/10.1016/j.nuclphysb.2015.06.023} {\bibfield
  {journal} {\bibinfo  {journal} {Nuclear Physics B}\ }\textbf {\bibinfo
  {volume} {898}},\ \bibinfo {pages} {113} (\bibinfo {year}
  {2015})}\BibitemShut {NoStop}%
\bibitem [{Note3()}]{Note3}%
  \BibitemOpen
  \bibinfo {note} {Note that $\omega ^2/\gamma \sim Q f $, the product of the
  oscillator's quality factor and frequency.}\BibitemShut {Stop}%
\bibitem [{\citenamefont {Susskind}(1994)}]{SusskindPRD94}%
  \BibitemOpen
  \bibfield  {author} {\bibinfo {author} {\bibfnamefont {L.}~\bibnamefont
  {Susskind}},\ }\href {\doibase 10.1103/PhysRevD.49.6606} {\bibfield
  {journal} {\bibinfo  {journal} {Phys. Rev. D}\ }\textbf {\bibinfo {volume}
  {49}},\ \bibinfo {pages} {6606} (\bibinfo {year} {1994})}\BibitemShut
  {NoStop}%
\end{thebibliography}%

\clearpage
\newpage

\begin{widetext}

\section*{Supplementary materials for ``Quantum gravitational decoherence of a mechanical oscillator from spacetime fluctuations''}


\section{Derivation of the master equation (\ref{finalME})}\label{si:MEderiv}

The starting point of our analysis is Eq. (\ref{ModDyn}) of the main
text. First, we redefined $\hat{H}$ and $\hat{H}_{\beta}$ in such a way
that the average of the perturbative part is zero. We can do this
by introducing 
\begin{equation}\label{h'}
\hat{H}'=\hat{H}+g(\hat{K})
\end{equation}
and
\begin{equation}
\hat{H}'_{\beta}(t)=\hat{H}_{\beta}(t)-g(\hat{K}),
\end{equation}
where
\begin{equation}\label{hb'}
g(\hat{K}):=\mathbb{E}[\hat{H}_{\beta}(t)]=4a_{P}\hat{K}^{2}\overline{\beta}.
\end{equation}

Following the standard approach, one first move to the interaction
picture with respect to $\hat{H}'$ i.e. $|\psi^{I}(t)\rangle:=e^{\frac{i}{\hbar}\hat{H}'t}|\psi(t)\rangle$
so that $\hat{\varrho}^{I}(t):=|\psi^{I}(t)\rangle\langle\psi^{I}(t)|$
satisfies the equation:
\begin{equation}\label{meint}
\partial_{t}\hat{\varrho}^{I}(t)=-\frac{i}{\hbar}\left[\hat{H'}_{\beta}^{I}(t),\hat{\varrho}^{I}(t)\right],
\end{equation}
which has the formal solution
\begin{equation}\label{formal_sol}
\hat{\varrho}^{I}(t)=\hat{\varrho}^{I}(0)-\frac{i}{\hbar}\int_{0}^{t}dt'\left[\hat{H'}_{\beta}^{I}(t'),\hat{\varrho}^{I}(t')\right].
\end{equation}
By replacing Eq. (\ref{formal_sol}) in Eq. (\ref{meint}) and using
the Markov approximation $\hat{\varrho}^{I}(t')\simeq\hat{\varrho}^{I}(t)$
one gets
\begin{equation}\label{mevarrho}
\partial_{t}{\hat{\varrho}}^{I}(t)=-\frac{i}{\hbar}\left[{\hat{H'}}_{\beta}^{I}(t),\hat{\varrho}^{I}(0)\right]-\frac{1}{\hbar^{2}}\int_{0}^{t}dt'\left[\hat{H'}_{\beta}^{I}(t),\left[\hat{H'}_{\beta}^{I}(t'),\hat{\varrho}^{I}(t)\right]\right]
\end{equation}
We now average over the noise both sides of the equation to get
the evolution equation of the statistical operator $\hat{\rho}^{I}(t)=\mathbb{E}[\hat{\varrho}^{I}(t)]$.
The first term in the RHS of Eq. (\ref{mevarrho}) averages to zero since $\hat{\varrho}^{I}(0)$
is not a random quantity and $\mathbb{E}[\hat{H'}_{\beta}^{I}(t)]=e^{i\hat{H}'t}\mathbb{E}[\hat{H}'_{\beta}(t)]e^{-i\hat{H}'t}=0$.   Regarding
the second term, using the analog of the Born approximation
when the system is  interacting with a quantum bath instead of a classical noise, one can factorize the noise average as $\mathbb{E}\left[\beta(t)\beta(t')\hat{\varrho}^{I}(t)\right]=\mathbb{E}[\beta(t)\beta(t')]\mathbb{E}\left[\hat{\varrho}^{I}(t)\right]$. Then, using Eq. \eqref{eq:BetaFl} (note that $\hat{H}'_{\beta}(t)\propto \beta(t)-\overline{\beta}$), Eq. (\ref{mevarrho}) becomes:  
\begin{equation}
\partial_{t}\hat{\rho}^{I}(t)=-\frac{16a_{P}^{2}\kappa}{\hbar^{2}}\int_{0}^{t}dt'f(t-t')\left[\hat{K}^{I2}(t),\left[\hat{K}^{I2}(t'),\hat{\rho}^{I}(t)\right]\right].\label{ME int}
\end{equation}
Moving back to the Schr\"odinger picture, after a little algebra one gets
\begin{equation}\label{Me finalsupp}
\partial_{t}\hat{\rho}(t)=-\frac{i}{\hbar}\left[\hat{H}+g(\hat{K}),\rho(t)\right]-\frac{16a_{P}^{2}\kappa}{\hbar^{2}}\int_{0}^{t}dt'f(t-t')\left[\hat{K}^{2},\left[\hat{K}^{I2}(t'-t),\hat{\rho}(t)\right]\right],
\end{equation}
which is precisely Eq.
(\ref{finalME}) in the main text. Note that if one considers from the start a white noise, the master equation can be derived exactly using It\^o calculus and one arrives at an equation of the form in Eq. (\ref{Me finalsupp}) with $f(t-t')=\delta(t-t')$.

For a free particle $\hat{H}=\hat{K}$, which implies $\hat{K}^{I2}(t'-t)=\hat{K}^{2}$
in Eq. (\ref{Me finalsupp}). Then the momentum eigenstates $|\boldsymbol{p}\rangle$
are eigenstates of the unitary part as well as of the Lindblad operators. By considering the matrix elements in the momentum basis $\hat{\rho}_{ab}(t):=\langle\boldsymbol{p}_{a}|\hat{\rho}(t)|\boldsymbol{p}_{b}\rangle$,
Eq. (\ref{Me finalsupp}) becomes: 
\begin{equation}\label{MEfreesupp}
\partial_{t}\hat{\rho}_{ab}(t)=\left\{ -\frac{i}{\hbar}\left(\Delta E_{1}+4a_{P}\Delta E_{2}\overline{\beta}\right)-\frac{16a_{P}^{2}\kappa}{\hbar^{2}}\int_{0}^{t}dt'f(t-t')(\Delta E_{2})^{2}\right\} \hat{\rho}_{ab}(t)
\end{equation}
where we introduced $\Delta E_{k}:=K_{a}^{k}-K_{b}^{k}=[(p^2_a/2m)^k-(p^2_b/2m)^k]$. The solution of Eq. (\ref{MEfreesupp}) is precisely Eq. (\ref{sol_free}) in the main text.
\\

The main difference in our derivation with respect to the one done in \cite{Petru21}, besides the fact that we consider possible non-white noises,  is that we performed the steps in Eqs. (\ref{meint}), (\ref{formal_sol}) and (\ref{mevarrho}) using $H'$ and $H_\beta'$ introduced in Eqs. (\ref{h'}) and (\ref{hb'})   instead of the original  $H$ and $H_\beta$. This is crucial for three reasons: (\textit{i}) to justify why the first term in Eq. (\ref{mevarrho}) averages to zero; (\textit{ii}) to get the correction to the free Hamiltonian $g(K)$ expected in the limit of no fluctuations ($\kappa=0$) and (\textit{iii}) to obtain the correct Lindbladian given the noises $\beta(t)$ defined as in Eq. (\ref{eq:BetaFl}) of the main text.

\section{Study of the coherences}\label{sec_coe}

Our starting point is the master equation (\ref{me_rwa_main2}) in the main text, which we report here for convenience:
\begin{equation}
\partial_{t}\hat{\rho}(t)=-\frac{i}{\hbar}\left[\hat{H}_{\text{RWA}},\hat{\rho}(t)\right]-\frac{16a_{P}^{2}\kappa}{\hbar^{2}}\int_{0}^{t}dt'f(t-t')\left[\hat{K}^{2},\left[\hat{K}^{I2}(t'-t),\hat{\rho}(t)\right]\right]
+\gamma \hat{a}\hat{\rho}(t)\hat{a}^{\dagger}-\frac{\gamma}{2}\left\{ \hat{N},\hat{\rho}(t)\right\} 
\label{Me324}
\end{equation}
with 
\begin{equation}
\hat{H}_{\text{RWA}}\equiv\hbar\omega(\hat{N}+\frac{1}{2})+a_{P}\overline{\beta}\frac{3\hbar^{2}\omega^{2}}{8}\left[\hat{N}^{2}+\hat{N}+\frac{1}{2}\right]\label{Hp324}
\end{equation}
and
\begin{equation}
\hat{K}^{2I}(t)\equiv e^{i\hat{H}_{\text{RWA}}t/\hbar}\hat{K}^{2}e^{-i\hat{H}_{\text{RWA}}t/\hbar},\label{KI324}
\end{equation}
where $\hat{K}=\frac{\hat{p}^{2}}{2m}.$

The goal is to find an approximate solution where only the second
term of Eq. (\ref{Me324}) is treated perturbatively. Note also that
while $\hat{H}_{\text{RWA}}$ is the Hamiltonian under the rotating wave approximation, we will
not apply the same approximation to the operators $\hat{K}^{2I}(t'-t)$
in the second term of Eq. (\ref{Me324}). This would certainly greatly simplify the analysis, but
we verified, by solving numerically exactly Eq. (\ref{Me324}),
that such an approximation leads to important errors in the evolution of $\hat{\rho(t)}$. 

We start by rewriting Eq. (\ref{Me324}) in terms of the super operators
$\mathcal{L}(t)=\mathcal{L}_{1}+\mathcal{L}_{2}(t)$
\begin{equation}
\partial_{t}\hat{\rho}(t)=\mathcal{L}(t)[\hat{\rho}(t)]=\mathcal{L}_{1}[\hat{\rho}(t)]+\mathcal{L}_{2}(t)[\hat{\rho}(t)]
\end{equation}
 where 
\begin{equation}
\mathcal{L}_{1}[\hat{\rho}(t)]=-\frac{i}{\hbar}\left[\hat{H}_{\text{RWA}},\hat{\rho}(t)\right]+\gamma \hat{a}\hat{\rho}(t)\hat{a}^{\dagger}-\frac{\gamma}{2}\left\{ \hat{N},\hat{\rho}(t)\right\} 
\end{equation}
with $\hat{N}=\hat{a}^\dagger\hat{a}$, and
\begin{equation}
\mathcal{L}_{2}(t)[\hat{\rho}(t)]=-\frac{16a_{P}^{2}\kappa}{\hbar^{2}}\int_{0}^{t}dt'f(t-t')\left[\hat{K}^{2},\left[\hat{K}^{I2}(t'-t),\hat{\rho}(t)\right]\right].
\end{equation}
The idea here is to describe the evolution due to $\mathcal{L}_{1}$ exactly while treating the one due to $\mathcal{L}_{2}(t)$ as a perturbation. It is convenient to introduce also the map $\Phi(t)$ that provides the evolution of
$\hat{\rho}(t)$ i.e. 
\begin{equation}
\hat{\rho}(t)=\Phi(t)[\hat{\rho}(0)],
\end{equation}
which fulfills the equation
\begin{equation}
\partial_{t}\Phi(t)=\mathcal{L}(t)[\Phi(t)]\;\;\;\;\;\;\textrm{with}\;\;\;\;\;\;\Phi(0)=\mathbb{1},\label{totmap324}
\end{equation}
as well as the state evolved only with $\mathcal{L}_{1}$ i.e.:
\begin{equation}
\hat{\rho}_1(t):=e^{\mathcal{L}_{1}t}[\hat{\rho}(0)].\label{varrhodef}
\end{equation}
The solution of Eq. (\ref{totmap324}) can be cast in the form
\begin{equation}
\Phi(t)=e^{\mathcal{L}_{1}t}\left(\mathbb{1}+\int_{0}^{t}d\tau e^{\mathcal{L}_{1}(-\tau)}\mathcal{L}_{2}(\tau)\Phi(\tau)\right),\label{sol324}
\end{equation}
which is written in a convenient form since we want to treat perturbatively only
$\mathcal{L}_{2}(\tau)$.

Given the exact solution in Eq. \eqref{sol324} and that we want to treat perturbatively the second term, we can approximate $\Phi(\tau)\simeq e^{\mathcal{L}_{1}\tau}$
inside the integral, obtaining: 
\begin{equation}
\Phi(t)\simeq e^{\mathcal{L}_{1}t}\left(\mathbb{1}+\int_{0}^{t}d\tau e^{\mathcal{L}_{1}(-\tau)}\mathcal{L}_{2}(\tau)e^{\mathcal{L}_{1}(\tau)}\right).\label{sol_app324}
\end{equation}
Since the second term is small, we further approximate the superoperators
$e^{\mathcal{L}_{1}\tau}$ inside the integral as 
\begin{equation}
e^{\mathcal{L}_{1}\tau}[...]\simeq e^{-\frac{i}{\hbar}\hat{H}_{\text{RWA}}\tau}...e^{+\frac{i}{\hbar}\hat{H}_{\text{RWA}}\tau},
\end{equation}
i.e. we neglect the amplitude damping in this term. 
Then
\begin{align}\label{40}
\hat{\rho}(t)&=\Phi(t)\hat{\rho}(0)\simeq e^{\mathcal{L}_{1}t}\left(\hat{\rho}(0)+\int_{0}^{t}d\tau e^{\frac{i}{\hbar}\hat{H}_{\text{RWA}}\tau}\mathcal{L}_{2}(\tau)\left[e^{-\frac{i}{\hbar}\hat{H}_{\text{RWA}}\tau}\hat{\rho}(0)e^{+\frac{i}{\hbar}\hat{H}_{\text{RWA}}\tau}\right]e^{-\frac{i}{\hbar}\hat{H}_{\text{RWA}}\tau}\right)=
\\
&=e^{\mathcal{L}_{1}t}\left(\rho(0)-\frac{16a_{P}^{2}\kappa}{\hbar^{2}}\int_{0}^{t}d\tau\int_{0}^{\tau}dt'f(\tau-t')e^{\frac{i}{\hbar}\hat{H}_{\text{RWA}}\tau}\left[\hat{K}^{2},\left[\hat{K}^{I2}(t'-\tau),e^{-\frac{i}{\hbar}\hat{H}_{\text{RWA}}\tau}\hat{\rho}(0)e^{+\frac{i}{\hbar}\hat{H}_{\text{RWA}}\tau}\right]\right]e^{-\frac{i}{\hbar}\hat{H}_{\text{RWA}}\tau}\right)\nonumber
\end{align}
The second term, after some manipulations, can be rewritten in the
simpler form, obtaining:
\begin{equation}
\hat{\rho}(t)\simeq e^{\mathcal{L}_{1}t}\left(\hat{\rho}(0)-\frac{16a_{P}^{2}\kappa}{\hbar^{2}}\int_{0}^{t}d\tau\int_{0}^{\tau}dt'f(\tau-t')\left[\hat{K}^{I2}(\tau),\left[\hat{K}^{I2}(t'),\hat{\rho}(0)\right]\right]\right).\label{finalresapp324}
\end{equation}
We are now interested in computing the matrix element $\langle0|\hat{\rho}(t)|1\rangle$,
which means we need to compute $\langle0|e^{\mathcal{L}_{1}t}\hat{\rho}(0)|1\rangle$
and $\langle0|e^{\mathcal{L}_{1}t}\left[\hat{K}^{2I}(\tau),\left[\hat{K}^{2I}(t'),\hat{\rho}(0)\right]\right]|1\rangle$. 
We will do it by taking as initial state:
\begin{equation}
\hat{\rho}(0)=\frac{|0\rangle+|1\rangle}{\sqrt{2}}\frac{\langle0|+\langle1|}{\sqrt{2}}=\frac{1}{2}\left(|0\rangle\langle0|+|1\rangle\langle0|+|0\rangle\langle1|+|1\rangle\langle1|\right).\label{r0324}
\end{equation}

\subsection{Computation of the term $\langle0|\hat{\rho}_1(t)|1\rangle=\langle0|e^{\mathcal{L}_{1}t}\rho(0)|1\rangle$}\label{sub_sec_01}

To compute this term we observe that $\hat{\rho}_1(t)$ defined in
Eq. (\ref{varrhodef}) by construction follows the master equation:
\begin{equation}
\frac{d \hat{\rho}_1(t)}{dt}=-\frac{i}{\hbar}\left[\hat{H}_{\text{RWA}},\hat{\rho}_1(t)\right]+\gamma\hat{a}\hat{\rho}_1(t)\hat{a}^{\dagger}-\frac{\gamma}{2}\left\{ \hat{N},\hat{\rho}_1(t)\right\} \label{ME_varrho}
\end{equation}
We take the matrix elements $\langle n_{1}|\hat{\rho}_1(t)|n_{2}\rangle$,
then Eq. (\ref{ME_varrho}) becomes:
\begin{equation}
\frac{d\langle n_{1}|\hat{\rho}_1(t)|n_{2}\rangle}{dt}=f(n_{1},n_{2},t)\langle n_{1}|\hat{\rho}_1(t)|n_{2}\rangle+\gamma\sqrt{(n_{1}+1)(n_{2}+1)}\langle n_{1}+1|\hat{\rho}_1(t)|n_{2}+1\rangle.\label{n1n2eq}
\end{equation}
where 
\begin{equation}
f(n_{1},n_{2},t)=-\frac{i}{\hbar}\left(E_{1}-E_{2}\right)-\frac{\gamma}{2}(n_{1}+n_{2})\label{fn1n2}
\end{equation}
with (see Eq. (\ref{Hp324}))
\begin{equation}\label{Ej}
E_{j}=\hbar\omega(n_{j}+\frac{1}{2})+a_{P}\overline{\beta}\frac{3\hbar^{2}\omega^{2}}{8}\left[n_{j}^{2}+n_{j}+\frac{1}{2}\right]
\end{equation}
for $j=1,2$. Eq.  (\ref{n1n2eq}) is of the form
\begin{equation}
\partial_{t}r(t)=f(t)r(t)+g(t),
\end{equation}
which has solution
\begin{equation}
r(t)=r(0)e^{\int_{0}^{t}f(t')dt'}+\int_{0}^{t}dse^{\int_{s}^{t}f(t')dt'}g(s)\,ds.
\end{equation}
This implies that the solution of Eq. (\ref{n1n2eq}) is
\begin{equation}
\langle n_{1}|\hat{\rho}_1(t)|n_{2}\rangle=\langle n_{1}|\hat{\rho}_1(0)|n_{2}\rangle e^{\int_{0}^{t}f(n_{1},n_{2},t')dt'}+\gamma\sqrt{(n_{1}+1)(n_{2}+1)}\int_{0}^{t}dse^{\int_{s}^{t}f(n_{1},n_{2},t')dt'}\langle n_{1}+1|\hat{\rho}_1(s)|n_{2}+1\rangle\label{solvarrho}
\end{equation}
One can find a more explicit solution by inserting iteratively Eq. (\ref{solvarrho})
in the matrix element inside the integral in the second term of Eq. (\ref{solvarrho}). As a result, one gets an infinite series of the form:
\begin{align}\label{series50}
\langle n_{1}|\hat{\rho}_1(t)|n_{2}\rangle&=\langle n_{1}|\hat{\rho}_1(0)|n_{2}\rangle e^{\int_{0}^{t}f(n_{1},n_{2},t')dt'}+
\\
&+\gamma\sqrt{(n_{1}+1)(n_{2}+1)}\int_{0}^{t}dse^{\int_{s}^{t}f(n_{1},n_{2},t')dt'}\langle n_{1}+1|\hat{\rho}_1(0)|n_{2}+1\rangle e^{\int_{0}^{s}f(n_{1}+1,n_{2}+1,t')dt'}+\nonumber
\\
&+\gamma^{2}\sqrt{(n_{1}+1)(n_{2}+1)(n_{1}+2)(n_{2}+2)}\int_{0}^{t}dse^{\int_{s}^{t}f(n_{1},n_{2},t')dt'}\int_{0}^{s}ds'e^{\int_{s'}^{s}f(n_{1}+1,n_{2}+1,t')dt'}\langle n_{1}+2|\hat{\rho}_1(0)|n_{2}+2\rangle+...\nonumber
\end{align}
Since we are interested in the case $n_{1}=0$ and $n_{2}=1$ and
the initial state is the one given in Eq. (\ref{r0324}), we have
\begin{align}
&\langle0|\hat{\rho}_1(t)|0\rangle=1-\frac{1}{2}e^{-\gamma t}
\\
&\langle0|\hat{\rho}_1(t)|1\rangle=\frac{1}{2}e^{-\frac{i}{\hbar}\left(E_{0}-E_{1}\right)t}e^{-\frac{\gamma}{2}t}\label{series-2}
\\
&\langle1|\hat{\rho}_1(t)|1\rangle=\frac{1}{2}e^{-\gamma t}\label{series-1}
\end{align}
where we used (see Eq. (\ref{fn1n2})) $f(0,0,t)=0$, $f(0,1,t)=-\frac{i}{\hbar}\left(E_{0}-E_{1}\right)-\frac{\gamma}{2}$
and $f(1,1,t)=-\gamma$.

\subsection{Computation of $\langle0|e^{\mathcal{L}_{1}t}\left[\hat{K}^{2I}(\tau),\left[\hat{K}^{2I}(t'),\hat{\rho}(0)\right]\right]|1\rangle$}

\subsubsection{General structure of the calculation}

Since this is the term that we want to compute perturbatively,  when applying the superoperator $e^{\mathcal{L}_{1}t}$ we stop to the first term of the series in Eq. \eqref{series50}.
Hence
\begin{equation}\label{54}
\langle0|e^{\mathcal{L}_{1}t}\left[\hat{K}^{2I}(\tau),\left[\hat{K}^{2I}(t'),\hat{\rho}(0)\right]\right]|1\rangle\simeq
e^{-\frac{i}{\hbar}\left(E_{0}-E_{1}\right)t}e^{-\frac{\gamma}{2}t}C(\tau,t')
\end{equation}
where we introduced
\begin{equation}
C(\tau,t'):=\langle0|\left[\hat{K}^{2I}(\tau),\left[\hat{K}^{2I}(t'),\hat{\rho}(0)\right]\right]|1\rangle=D_{1}(\tau,t')-D_{2}(\tau,t')-D_{2}(t',\tau)+D_{3}(\tau,t'),\label{Cdef243}
\end{equation}
with $\hat{K}^{2I}(\tau)$ given in Eq. (\ref{KI324}), $\hat{\rho}(0)$ in Eq.
(\ref{r0324}) and
\begin{align}
&D_{1}(\tau,t'):=\langle0|\hat{K}^{2I}(\tau)\hat{K}^{2I}(t')\hat{\rho}(0)|1\rangle;
\\
&D_{2}(\tau,t'):=\langle0|\hat{K}^{2I}(\tau)\hat{\rho}(0)\hat{K}^{2I}(t')|1\rangle;
\\
&D_{3}(\tau,t'):=\langle0|\hat{\rho}(0)\hat{K}^{2I}(t')\hat{K}^{2I}(\tau)|1\rangle.
\end{align}
Our goal is now to compute $D_1$, $D_2$ and $D_3$. This requires the calculation of several matrix elements for which it is convenient to rewrite $\hat{K}$ in terms of the ladder operators i.e. $\hat{K}=\frac{\hat{p}^{2}}{2m}=\frac{\hbar\omega}{4}\left[2\hat{N}+1-(\hat{a}^{\dagger})^{2}-\hat{a}^{2}\right]$.

We start by computing 
\begin{align}
\langle0|\hat{K}^{2I}(\tau)|0\rangle&=\langle0|\hat{K}^{2}|0\rangle=\sum_{n=0}^{\infty}\langle0|\frac{\hbar\omega}{4}\left[2\hat{N}+1-(\hat{a}^{\dagger})^{2}-\hat{a}^{2}\right]|n\rangle\langle n|\frac{\hbar\omega}{4}\left[2\hat{N}+1-(\hat{a}^{\dagger})^{2}-\hat{a}^{2}\right]|0\rangle=
\\
&=\left(\frac{\hbar\omega}{4}\right)^{2}\langle0|\left[2\hat{N}+1-(\hat{a}^{\dagger})^{2}-\hat{a}^{2}\right]|0\rangle\langle0|\left[2\hat{N}+1-(\hat{a}^{\dagger})^{2}-\hat{a}^{2}\right]|0\rangle+\nonumber
\\
&+\left(\frac{\hbar\omega}{4}\right)^{2}\langle0|\left[2\hat{N}+1-(\hat{a}^{\dagger})^{2}-\hat{a}^{2}\right]|2\rangle\langle2|\left[2\hat{N}+1-(\hat{a}^{\dagger})^{2}-\hat{a}^{2}\right]|0\rangle=\nonumber
\\
&=\left(\frac{\hbar\omega}{4}\right)^{2}+\left(\frac{\hbar\omega}{4}\right)^{2}\langle0|\hat{a}^{2}|2\rangle\langle2|(\hat{a}^{\dagger})^{2}|0\rangle=\left(\frac{\hbar\omega}{4}\right)^{2}+\left(\frac{\hbar\omega}{4}\right)^{2}2=\frac{3\hbar^{2}\omega^{2}}{16}.\nonumber  
\end{align}
When going from the first line to the second one, we used the fact that, given the form of the
operator $\hat{K}$, only the matrix elements between two number states
that are the same or differs by 2 survives. 

By performing analogues calculation, we can find these other matrix elements
(we also report $\langle0|K^{2I}(\tau)|0\rangle$ for convenience)
\begin{equation}
\langle0|\hat{K}^{2I}(\tau)|0\rangle=\frac{3\hbar^{2}\omega^{2}}{16},\label{00}
\end{equation}
\begin{equation}
\langle1|\hat{K}^{2I}(\tau)|1\rangle=\frac{15\hbar^{2}\omega^{2}}{16},\label{11}
\end{equation}
\begin{equation}
\langle0|\hat{K}^{2I}(\tau)|2\rangle=-\frac{3\sqrt{2}\hbar^{2}\omega^{2}}{8}e^{\frac{i}{\hbar}(E_{0}-E_{2})\tau},\label{02}
\end{equation}
\begin{equation}
\langle0|\hat{K}^{2I}(\tau)|4\rangle=\frac{\sqrt{6}\hbar^{2}\omega^{2}}{8}e^{\frac{i}{\hbar}(E_{0}-E_{4})\tau},\label{04}
\end{equation}
\begin{equation}
\langle1|\hat{K}^{2I}(\tau)|3\rangle=-\frac{5\sqrt{6}\hbar^{2}\omega^{2}}{8}e^{\frac{i}{\hbar}(E_{1}-E_{3})\tau},\label{13}
\end{equation}
\begin{equation}
\langle1|\hat{K}^{2I}(\tau)|5\rangle=\frac{\sqrt{30}\hbar^{2}\omega^{2}}{8}e^{\frac{i}{\hbar}(E_{1}-E_{5})\tau}.\label{15}
\end{equation}
where the energies $E_j$ are defined in Eq. \eqref{Ej}. 

\subsubsection{Computation of $D_{1}$}

We can now compute  
\begin{equation}
D_{1}(\tau,t')=\langle0|\hat{K}^{2I}(\tau)\hat{K}^{2I}(t')\hat{\rho}(0)|1\rangle=\frac{1}{2}\left(\langle0|\hat{K}^{2I}(\tau)\hat{K}^{2I}(t')|0\rangle+\langle0|\hat{K}^{2I}(\tau)\hat{K}^{2I}(t')|1\rangle\right)
\end{equation}
Since the operators $\hat{K}^{2I}(t')$ contains even number of ladder operators,
all matrix elements between an odd and an even state are zero, therefore
$\langle0|\hat{K}^{2I}(t')\hat{K}^{2I}(t'')|1\rangle=0$.

We are left with computing:
\begin{align}\label{28342}
D_{1}(\tau,t')&=\frac{1}{2}\langle0|\hat{K}^{2I}(\tau)\hat{K}^{2I}(t')|0\rangle=\frac{1}{2}\sum_{n=0}^{\infty}\langle0|\hat{K}^{2I}(\tau)|n\rangle\langle n|\hat{K}^{2I}(t')|0\rangle=
\\
&=\frac{1}{2}\langle0|\hat{K}^{2I}(\tau)|0\rangle\langle0|\hat{K}^{2I}(t')|0\rangle+\frac{1}{2}\langle0|\hat{K}^{2I}(\tau)|2\rangle\langle2|\hat{K}^{2I}(t')|0\rangle+\frac{1}{2}\langle0|\hat{K}^{2I}(\tau)|4\rangle\langle4|\hat{K}^{2I}(t')|0\rangle.\nonumber
\end{align}
The reason for the last step is that $\hat{K}^{2I}(t)$ contains at most
terms with the 4-th powers of the ladder operators (proportional to
$\hat{a}^{4}$ or $(\hat{a}^{\dagger})^{4})$ so at most can connect number states
that differs by 4 or less. The terms $n=1,3$ are zero once again because matrix elements between odd and even number states are always zero. 

By using the results in Eqs. (\ref{00}), (\ref{02}), and (\ref{04})
we get:
\begin{equation}
D_{1}(\tau,t')=\frac{9\hbar^{4}\omega^{4}}{512}+\frac{9\hbar^{4}\omega^{4}}{64}e^{\frac{i}{\hbar}(E_{0}-E_{2})(\tau-t')}+\frac{3\hbar^{4}\omega^{4}}{64}e^{\frac{i}{\hbar}(E_{0}-E_{4})(\tau-t')}.
\end{equation}

\subsubsection{Computation of $D_{2}$}

We proceed by computing:
\begin{align}
D_{2}(\tau,t')&=\langle0|\hat{K}^{2I}(\tau)\hat{\rho}(0)\hat{K}^{2I}(t')|1\rangle=\frac{1}{2}\left\{ \langle0|\hat{K}^{2I}(\tau)|0\rangle\langle0|\hat{K}^{2I}(t')|1\rangle+\langle0|\hat{K}^{2I}(\tau)|0\rangle\langle1|\hat{K}^{2I}(t')|1\rangle\right.+
\\
&\left.+\langle0|\hat{K}^{2I}(\tau)|1\rangle\langle0|\hat{K}^{2I}(t')|1\rangle+\langle0|\hat{K}^{2I}(\tau)|1\rangle\langle1|\hat{K}^{2I}(t')|1\rangle\right\} =\frac{\left(\langle0|\hat{K}^{2I}(\tau)|0\rangle+\langle0|\hat{K}^{2I}(\tau)|1\rangle\right)\left(\langle0|\hat{K}^{2I}(t')|1\rangle+\langle1|\hat{K}^{2I}(t')|1\rangle\right)}{2} \nonumber   
\end{align}
once again the terms $\langle0|\hat{K}^{2I}(\tau)|1\rangle=\langle0|\hat{K}^{2I}(t')|1\rangle=0$
and using Eqs. (\ref{00}) and (\ref{11}) we get
\begin{equation}
D_{2}(\tau,t')=\frac{\left(\frac{3\hbar^{2}\omega^{2}}{16}\right)\left(\frac{15\hbar^{2}\omega^{2}}{16}\right)}{2}=\frac{45\hbar^{4}\omega^{4}}{512}
\end{equation}

\subsubsection{Computation of $D_{3}$}
We finally compute
\begin{align}\label{71}
D_{3}(\tau,t')&=\langle0|\hat{\rho}(0)\hat{K}^{2I}(t')\hat{K}^{2I}(\tau)|1\rangle=\frac{1}{2}\left(\langle0|\hat{K}^{2I}(t')\hat{K}^{2I}(\tau)|1\rangle+\langle1|\hat{K}^{2I}(t')\hat{K}^{2I}(\tau)|1\rangle\right)
\\
&=\frac{1}{2}\langle1|\hat{K}^{2I}(t')\hat{K}^{2I}(\tau)|1\rangle=\frac{1}{2}\sum_{n=0}^{\infty}\langle1|\hat{K}^{2I}(t')|n\rangle\langle n|\hat{K}^{2I}(\tau)|1\rangle=\nonumber
\\
&=\frac{1}{2}\left(\langle1|\hat{K}^{2I}(t')|1\rangle\langle1|\hat{K}^{2I}(\tau)|1\rangle+\langle1|\hat{K}^{2I}(t')|3\rangle\langle3|\hat{K}^{2I}(\tau)|1\rangle+\langle1|\hat{K}^{2I}(t')|5\rangle\langle5|\hat{K}^{2I}(\tau)|1\rangle\right).\nonumber
\end{align}
Using Eqs. (\ref{11}), (\ref{13}) and (\ref{15}) we get 
\begin{equation}
D_{3}(\tau,t')=\left(\frac{225\hbar^{4}\omega^{4}}{512}\right)+\frac{75\hbar^{4}\omega^{4}}{64}e^{\frac{i}{\hbar}(E_{1}-E_{3})(t'-\tau)}+\frac{15\hbar^{4}\omega^{4}}{64}e^{\frac{i}{\hbar}(E_{1}-E_{5})(t'-\tau)}.
\end{equation}

\subsubsection{Computation of $C(\tau,t')$}
In summary we found 
\begin{align}
D_{1}(\tau,t')&=\frac{9\hbar^{4}\omega^{4}}{512}+\frac{9\hbar^{4}\omega^{4}}{64}e^{\frac{i}{\hbar}(E_{0}-E_{2})(\tau-t')}+\frac{3\hbar^{4}\omega^{4}}{64}e^{\frac{i}{\hbar}(E_{0}-E_{4})(\tau-t')}
\\
D_{2}(\tau,t')&=\frac{45\hbar^{4}\omega^{4}}{512}
\\
D_{3}(\tau,t')&=\frac{225\hbar^{4}\omega^{4}}{512}+\frac{75\hbar^{4}\omega^{4}}{64}e^{\frac{i}{\hbar}(E_{1}-E_{3})(t'-\tau)}+\frac{15\hbar^{4}\omega^{4}}{64}e^{\frac{i}{\hbar}(E_{1}-E_{5})(t'-\tau)}.
\end{align}
Then, going back to Eq. (\ref{Cdef243}), we can finally find:
\begin{equation}
C(\tau,t')=\frac{9\hbar^{4}\omega^{4}}{32}+\frac{9\hbar^{4}\omega^{4}}{64}e^{\frac{i}{\hbar}(E_{0}-E_{2})(\tau-t')}+\frac{3\hbar^{4}\omega^{4}}{64}e^{\frac{i}{\hbar}(E_{0}-E_{4})(\tau-t')}+\frac{75\hbar^{4}\omega^{4}}{64}e^{\frac{i}{\hbar}(E_{1}-E_{3})(t'-\tau)}+\frac{15\hbar^{4}\omega^{4}}{64}e^{\frac{i}{\hbar}(E_{1}-E_{5})(t'-\tau)}\label{finalC324}
\end{equation}

\subsection{Final results}
By taking the matrix element $\langle0|...|1\rangle$ in Eq. (\ref{finalresapp324})
we find:
\begin{align}
\langle0|\hat{\rho}(t)|1\rangle&=\langle0|\hat{\rho}_{1}(t)|1\rangle-\frac{16a_{P}^{2}\kappa}{\hbar^{2}}\int_{0}^{t}d\tau\int_{0}^{\tau}dt'f(\tau-t')\langle0|e^{\mathcal{L}_{1}t}\left[\hat{K}^{I2}(\tau),\left[\hat{K}^{I2}(t'),\hat{\rho}(0)\right]\right]|1\rangle\simeq
\\
&\simeq\frac{1}{2}e^{-\frac{i}{\hbar}\left(E_{0}-E_{1}\right)t}e^{-\frac{\gamma}{2}t}\left(1-\frac{32a_{P}^{2}\kappa}{\hbar^{2}}\int_{0}^{t}d\tau\int_{0}^{\tau}dt'f(\tau-t')C(\tau,t')\right)
\end{align}
with $\langle0|\hat{\rho}_1(t)|1\rangle$ given in Eq. (\ref{series-2})
and we used Eqs. \eqref{54} and \eqref{Cdef243}, with $C(\tau,t')$ given in Eq. (\ref{finalC324}).

In the white noise limit i.e. $f(\tau-t')=\delta(\tau-t')$, Eq. (\ref{finalC324}) reduces to $C(\tau,\tau)=\frac{15}{8}\hbar^{4}\omega^{4}$ and one gets 
\begin{equation}\label{ro01}
\langle0|\hat{\rho}(t)|1\rangle\simeq\frac{1}{2}e^{-\frac{i}{\hbar}\left(E_{0}-E_{1}\right)t}e^{-\frac{\gamma}{2}t}\left(1-30\hbar^{2}\omega^{4}a_{P}^{2}\kappa t\right),
\end{equation}
which is Eq. \eqref{r01final} of the main text.

\section{Study of the decay of the excited state}
In this section we study
the decay of the excited state. Precisely, we prepare the system in
the state $\hat{\rho}(0)=|1\rangle\langle1|$ and study how it evolves. 

Starting from Eq. \eqref{finalresapp324} and by taking the matrix element $\langle1|...|1\rangle$
one gets: 
\begin{equation}
\langle1|\hat{\rho}(t)|1\rangle=\langle1|\hat{\rho}_{1}(t)|1\rangle-\frac{16a_{P}^{2}\kappa}{\hbar^{2}}\int_{0}^{t}d\tau\int_{0}^{\tau}dt'f(\tau-t')\langle1|e^{\mathcal{L}_{1}t}\left[\hat{K}^{I2}(\tau),\left[\hat{K}^{I2}(t'),\hat{\rho}(0)\right]\right]|1\rangle.\label{mat_el_11}
\end{equation}
The first term gives the standard amplitude damping $\langle1|\hat{\rho}_{1}(t)|1\rangle=e^{-\gamma t}$,
while the second one, as in the previous section, can be approximated
as: 
\begin{equation}
\langle1|e^{\mathcal{L}_{1}t}\left[\hat{K}^{I2}(\tau),\left[\hat{K}^{I2}(t'),\hat{\rho}(0)\right]\right]|1\rangle\simeq e^{-\gamma t}\langle1|\left[\hat{K}^{I2}(\tau),\left[\hat{K}^{I2}(t'),\hat{\rho}(0)\right]\right]|1\rangle.
\end{equation}
We focus on computing 
\begin{equation}\label{82}
\langle1|\left[\hat{K}^{I2}(\tau),\left[\hat{K}^{I2}(t'),\hat{\rho}(0)\right]\right]|1\rangle=D_{1}^{(1)}(\tau,t')-D_{2}^{(1)}(\tau,t')-D_{2}^{(1)}(t',\tau)+D_{3}^{(1)}(\tau,t')
\end{equation}
with
\begin{align}
D_{1}^{(1)}(\tau,t')&:=\langle1|\hat{K}^{I2}(\tau)\hat{K}^{I2}(t')\hat{\rho}(0)|1\rangle=\langle1|\hat{K}^{I2}(\tau)\hat{K}^{I2}(t')|1\rangle=\textrm{see Eq. (\ref{71})}=2D_{3}(t',\tau)=
\\
&=\frac{225\hbar^{4}\omega^{4}}{256}+\frac{75\hbar^{4}\omega^{4}}{32}e^{-\frac{i}{\hbar}(E_{1}-E_{3})(t'-\tau)}+\frac{15\hbar^{4}\omega^{4}}{32}e^{-\frac{i}{\hbar}(E_{1}-E_{5})(t'-\tau)};
\\
D_{2}^{(1)}(\tau,t')&:=\langle1|\hat{K}^{I2}(\tau)\hat{\rho}(0)\hat{K}^{I2}(t')|1\rangle=\langle1|\hat{K}^{I2}(\tau)|1\rangle\langle1|\hat{K}^{I2}(t')|1\rangle=\left(\frac{15\hbar^{2}\omega^{2}}{16}\right)^{2}=\frac{225\hbar^{4}\omega^{4}}{256};
\\
D_{3}^{(1)}(\tau,t')&:=\langle1|\hat{\rho}(0)\hat{K}^{I2}(t')\hat{K}^{I2}(\tau)|1\rangle=\langle1|\hat{K}^{2I}(t')\hat{K}^{2I}(\tau)|1\rangle=\textrm{see Eq. (\ref{71})}=2D_{3}(\tau,t')=
\\
&=\frac{225\hbar^{4}\omega^{4}}{256}+\frac{75\hbar^{4}\omega^{4}}{32}e^{\frac{i}{\hbar}(E_{1}-E_{3})(t'-\tau)}+\frac{15\hbar^{4}\omega^{4}}{32}e^{\frac{i}{\hbar}(E_{1}-E_{5})(t'-\tau)}.
\end{align}

In the white noise limit ($f(\tau-t')=\delta(\tau-t')$), Eq. (\ref{82}) simplifies to
\begin{equation}
\langle1|\left[\hat{K}^{I2}(\tau),\left[\hat{K}^{I2}(t'),\hat{\rho}(0)\right]\right]|1\rangle=\frac{45\hbar^{4}\omega^{4}}{8}
\end{equation}
and Eq. (\ref{mat_el_11}) becomes
\begin{equation}\label{ro11_sm}
\langle1|\hat{\rho}(t)|1\rangle=e^{-\gamma t}\left(1-45\hbar^{2}\omega^{4}a_{P}^{2}\kappa t\right),
\end{equation}
which is Eq. (\ref{overlaps}) of the main text. 

\section{Study of the evolution of the ground state}

In this section we study the evolution of the ground state. Hence
we prepare the system in the ground state $\hat{\rho}(0)=|0\rangle\langle0|$
and study how this evolve. 

Once again, starting from Eq. \eqref{finalresapp324} and by taking the matrix element
$\langle0|...|0\rangle$ one gets: 

\begin{equation}
\langle0|\hat{\rho}(t)|0\rangle=\langle0|\hat{\rho}_{1}(t)|0\rangle-\frac{16a_{P}^{2}\kappa}{\hbar^{2}}\int_{0}^{t}d\tau\int_{0}^{\tau}dt'f(\tau-t')\langle0|e^{\mathcal{L}_{1}t}\left[\hat{K}^{I2}(\tau),\left[\hat{K}^{I2}(t'),\hat{\rho}(0)\right]\right]|0\rangle.\label{mat_el_00}
\end{equation}
The first term is simply $\langle0|\hat{\rho}_{1}(t)|0\rangle=1$,
since the ground state is stable with respect to the amplitude damping.
The second term can be approximated as 
\begin{equation}
\langle0|e^{\mathcal{L}_{1}t}\left[\hat{K}^{I2}(\tau),\left[\hat{K}^{I2}(t'),\hat{\rho}(0)\right]\right]|0\rangle\simeq\langle0|\left[\hat{K}^{I2}(\tau),\left[\hat{K}^{I2}(t'),\hat{\rho}(0)\right]\right]|0\rangle.
\end{equation}
We focus on 
\begin{equation}
\langle0|\left[\hat{K}^{I2}(\tau),\left[\hat{K}^{I2}(t'),\hat{\rho}(0)\right]\right]|0\rangle=D_{1}^{(0)}(\tau,t')-D_{2}^{(0)}(\tau,t')-D_{2}^{(0)}(t',\tau)+D_{3}^{(0)}(\tau,t')
\end{equation}
with
\begin{align}
D_{1}^{(0)}(\tau,t')&:=\langle0|\hat{K}^{I2}(\tau)\hat{K}^{I2}(t')\hat{\rho}(0)|0\rangle=\langle0|\hat{K}^{I2}(\tau)\hat{K}^{I2}(t')|0\rangle=\textrm{see Eq. (\ref{28342})}=2D_{1}(\tau,t')=
\\
&=\frac{9\hbar^{4}\omega^{4}}{256}+\frac{9\hbar^{4}\omega^{4}}{32}e^{\frac{i}{\hbar}(E_{0}-E_{2})(\tau-t')}+\frac{3\hbar^{4}\omega^{4}}{32}e^{\frac{i}{\hbar}(E_{0}-E_{4})(\tau-t')};\nonumber
\\
D_{2}^{(0)}(\tau,t')&:=\langle0|\hat{K}^{I2}(\tau)\hat{\rho}(0)\hat{K}^{I2}(t')|0\rangle=\langle0|\hat{K}^{I2}(\tau)|0\rangle\langle0|\hat{K}^{I2}(t')|0\rangle=\left(\frac{3\hbar^{2}\omega^{2}}{16}\right)^{2}=\frac{9\hbar^{4}\omega^{4}}{256};
\\
D_{3}^{(0)}(\tau,t')&:=\langle0|\hat{\rho}(0)\hat{K}^{I2}(t')\hat{K}^{I2}(\tau)|0\rangle=\langle0|\hat{K}^{2I}(t')\hat{K}^{2I}(\tau)|0\rangle=\textrm{see Eq. (\ref{28342})}=2D_{1}(t',\tau)=
\\
&=\frac{9\hbar^{4}\omega^{4}}{256}+\frac{9\hbar^{4}\omega^{4}}{32}e^{-\frac{i}{\hbar}(E_{0}-E_{2})(\tau-t')}+\frac{3\hbar^{4}\omega^{4}}{32}e^{-\frac{i}{\hbar}(E_{0}-E_{4})(\tau-t')}.\nonumber
\end{align}
In the white noise limit ($f(\tau-t')=\delta(\tau-t')$), the matrix
element simplifies to
\begin{equation}
\langle0|\left[\hat{K}^{I2}(\tau),\left[\hat{K}^{I2}(t'),\hat{\rho}(0)\right]\right]|0\rangle=\frac{3\hbar^{4}\omega^{4}}{4}
\end{equation}
and Eq. (\ref{mat_el_00}) becomes
\begin{equation}\label{ro00_sm}
\langle0|\hat{\rho}(t)|0\rangle=1-6\hbar^{2}\omega^{4}a_{P}^{2}\kappa t,
\end{equation}
which is Eq.~(\ref{overlaps}) of the main text.


\clearpage
\newpage

\section{Calculations for the metric fluctuations model}

The Markovian limit of the model proposed in \cite{breuer2009metric} is
\begin{equation}
i\hbar\frac{d}{dt}|\psi\rangle=\left[\hat{H}+\hat{H}_{p}\right]|\psi\rangle
\end{equation}
with 
\begin{equation}
\hat{H}_{p}= w(t) \sqrt{\tau_{c}} \hat{K} 
\end{equation}
and $w(t)$ being a white noise i.e. 
\begin{equation}
\langle w(t)\rangle=0\;\;\;\;\;\;\; \langle w(t)w(t') \rangle=\delta(t-t') \;.
\end{equation}
The corresponding master equation is given by \cite{breuer2009metric}
\begin{equation}
\frac{d\hat{\rho}(t)}{dt}=-\frac{i}{\hbar}\left[\hat{H},\hat{\rho}(t)\right]-\frac{\tau_{c}}{2\hbar^{2}}\left[\left[\hat{K},\left[\hat{K},\hat{\rho}(t)\right]\right]\right] \;.
\end{equation}
When considering a free particle $\hat{H}=\hat{K}$,
the evolution of the matrix elements in momentum basis $\hat{\rho}_{ab}(t):=\langle\boldsymbol{p}_{a}|\hat{\rho}(t)|\boldsymbol{p}_{b}\rangle$ is
\begin{equation}
\hat{\rho}_{ab}(t)=\hat{\rho}_{ab}(0)e^{-\frac{i}{\hbar}\Delta E_{ab}t}e^{-\frac{\tau_{c}}{2\hbar^{2}}\Delta E_{ab}^{2}t} \;,
\end{equation}
with $\Delta E_{ab}=[(\boldsymbol{p}_{a}^{2}/2m)-(\boldsymbol{p}_{b}^{2}/2m)]$.

\subsection{The harmonic oscillator}

Similarly to what we did for the model with fluctuating minimal length, we now focus on the dynamics in one dimension of an harmonic
oscillator including amplitude damping. 
Then the Master equation becomes
\begin{equation}\label{ME_breu_osci}
\partial_{t}\hat{\rho}(t)=-\frac{i}{\hbar}\left[\hat{H}_{0},\hat{\rho}(t)\right]-\frac{\tau_{c}}{2\hbar^{2}}\left[\left[\hat{K},\left[\hat{K},\hat{\rho}(t)\right]\right]\right]+\gamma a\hat{\rho}(t)a^{\dagger}-\frac{\gamma}{2}\left\{ \hat{N},\hat{\rho}(t)\right\} 
\end{equation}
with $\hat{H}_{0}=\hbar\omega\hat{N}$ with $\omega$ the frequency of
the oscillator and $\hat{K}=\hat{p}^{2}/2m$ as before. Note that,
since in this model there are no modifications of the unitary term, the Hamiltonian 
does not require any rotating wave approximation. 

As for the previous model, we split the RHS term of Eq. (\ref{ME_breu_osci}) in two
terms:
\begin{equation}\label{L1_breu}
\mathcal{L}_{1}[\hat{\rho}(t)]=-\frac{i}{\hbar}\left[\hat{H}_{0},\hat{\rho}(t)\right]+\gamma\hat{a}\hat{\rho}(t)\hat{a}^{\dagger}-\frac{\gamma}{2}\left\{ \hat{N},\hat{\rho}(t)\right\},
\end{equation}
which we solve exactly and 
\begin{equation}\label{L2_Breu}
\mathcal{L}_{2}[\hat{\rho}(t)]=-\frac{\tau_{c}}{2\hbar^{2}}\left[\hat{K},\left[\hat{K},\hat{\rho}(t)\right]\right],
\end{equation}
which we treat perturbatively. 

Following the analysis done in section \ref{sec_coe}, the approximate solution of Eq. (\ref{ME_breu_osci}) is (see in
particular Eq. \eqref{40}):
\begin{align}\label{App_sol_breu}
\hat{\rho}(t)&\simeq e^{\mathcal{L}_{1}t}\left(\hat{\rho}(0)+\int_{0}^{t}d\tau e^{\frac{i}{\hbar}\hat{H}_{0}\tau}\mathcal{L}_{2}\left[e^{-\frac{i}{\hbar}\hat{H}_{0}\tau}\hat{\rho}(0)e^{+\frac{i}{\hbar}\hat{H}_{0}\tau}\right]e^{-\frac{i}{\hbar}\hat{H}_{0}\tau}\right)\nonumber
\\
&=e^{\mathcal{L}_{1}t}\left(\hat{\rho}(0)-\frac{\tau_{c}}{2\hbar^{2}}\int_{0}^{t}d\tau e^{\frac{i}{\hbar}\hat{H}_{0}\tau}\left[\hat{K},\left[\hat{K},e^{-\frac{i}{\hbar}\hat{H}_{0}\tau}\hat{\rho}(0)e^{+\frac{i}{\hbar}\hat{H}_{0}\tau}\right]\right]e^{-\frac{i}{\hbar}\hat{H}_{0}\tau}\right)=\nonumber
\\
&=e^{\mathcal{L}_{1}t}\left(\hat{\rho}(0)-\frac{\tau_{c}}{2\hbar^{2}}\int_{0}^{t}d\tau\left[\hat{K}^{I}(\tau),\left[\hat{K}^{I}(\tau),\hat{\rho}(0)\right]\right]\right),
\end{align}
with 
\begin{equation}
\hat{K}^{I}(\tau)=e^{\frac{i}{\hbar}\hat{H}_{0}\tau}\hat{K}e^{-\frac{i}{\hbar}\hat{H}_{0}\tau}.
\end{equation}

\subsection{Study of the coherences}

As before, we consider the situation where the oscillator is
prepared in the state $|0\rangle+|1\rangle/\sqrt{2}$ i. e. 
\begin{equation}
\hat{\rho}(0)=\frac{1}{2}\left(|0\rangle\langle0|+|1\rangle\langle0|+|0\rangle\langle1|+|1\rangle\langle1|\right)
\end{equation}
and we study the time evolution of the matrix element $\langle0|\hat{\rho}(t)|1\rangle$.
Therefore, given Eq. (\ref{App_sol_breu}), we need to compute $\langle0|e^{\mathcal{L}_{1}t}\hat{\rho}(0)|1\rangle$
and $\langle0|e^{\mathcal{L}_{1}t}\left[\hat{K}^{I}(\tau),\left[\hat{K}^{I}(\tau),\hat{\rho}(0)\right]\right]|1\rangle$. 

\subsection{Computation of $\langle0|e^{\mathcal{L}_{1}t}\hat{\rho}(0)|1\rangle$}

This is precisely the same calculation as the one done in subsection \ref{sec_coe}, with the only difference that the energies eigenstates are
not given anymore by Eq. \eqref{Ej} but simply by $E_{j}=\hbar\omega n_{j}$.
Then we have (see Eq. (\ref{series-2})):
\begin{equation}
\langle0|e^{\mathcal{L}_{1}t}\hat{\rho}(0)|1\rangle=\frac{1}{2}e^{i\omega t}e^{-\frac{\gamma}{2}t}.\label{bobo1}
\end{equation}

\subsection{Computation of $\langle0|e^{\mathcal{L}_{1}t}\left[\hat{K}^{I}(\tau),\left[\hat{K}^{I}(\tau),\hat{\rho}(0)\right]\right]|1\rangle$}

As done in section \ref{sec_coe}, we approximate 
\begin{equation}
\langle0|e^{\mathcal{L}_{1}t}\left[\hat{K}^{I}(\tau),\left[\hat{K}^{I}(\tau),\hat{\rho}(0)\right]\right]|1\rangle\simeq e^{i\omega t}e^{-\frac{\gamma}{2}t}B(\tau)
\end{equation}
where 
\begin{equation}
B(\tau):=\langle0|\left[\hat{K}^{I}(\tau),\left[\hat{K}^{I}(\tau),\hat{\rho}(0)\right]\right]|1\rangle=D_{1}(\tau)-2D_{2}(\tau)+D_{3}(\tau)
\end{equation}
with \begin{align}
D_{1}(\tau)&:=\langle0|\hat{K}^{I2}(\tau)\hat{\rho}(0)|1\rangle\label{D1_bre}
\\
D_{2}(\tau)&:=\langle0|\hat{K}^{I}(\tau)\hat{\rho}(0)\hat{K}^{I}(\tau)|1\rangle\label{D2_bre}
\\
D_{3}(\tau)&:=\langle0|\hat{\rho}(0)\hat{K}^{I2}(\tau)|1\rangle\label{D3_bre}
\end{align}
These three contributions can be easily computed:
\begin{equation}
D_{1}(\tau)=\frac{1}{2}\left(\langle0|\hat{K}^{I2}(\tau)|0\rangle+\langle0|\hat{K}^{I2}(\tau)|1\rangle\right)=\frac{1}{2}\langle0|\hat{K}^{I2}(\tau)|0\rangle=\frac{3\hbar^{2}\omega^{2}}{32},
\end{equation}
where we used Eq. (\ref{00}) and the fact that $\langle n_{1}|\hat{K}^{I2}(\tau)|n_{2}\rangle=0$
when $n_{2}-n_{1}$ is an odd number. 

Similarly, using Eq. (\ref{11}), we can compute
\begin{equation}
D_{3}(\tau)=\frac{1}{2}\left(\langle0|\hat{K}^{I2}(\tau)|1\rangle+\langle1|\hat{K}^{I2}(\tau)|1\rangle\right)=\frac{1}{2}\langle1|\hat{K}^{I2}(\tau)|1\rangle=\frac{15\hbar^{2}\omega^{2}}{32}.
\end{equation}
Finally we compute
\begin{equation}
D_{2}(\tau)=\frac{1}{2}\left(\langle0|\hat{K}^{I}(\tau)|0\rangle\langle0|\hat{K}^{I}(\tau)|1\rangle+\langle0|\hat{K}^{I}(\tau)|1\rangle\langle0|\hat{K}^{I}(\tau)|1\rangle\right.+\left.+\langle0|\hat{K}^{I}(\tau)|0\rangle\langle1|\hat{K}^{I}(\tau)|1\rangle+\langle0|\hat{K}^{I}(\tau)|1\rangle\langle1|\hat{K}^{I}(\tau)|1\rangle\right).
\end{equation}
Since $\hat{K}^{I}(\tau)=\frac{\hbar\omega}{4}\left[2\hat{N}+1-(\hat{a}^{\dagger})^{2}e^{+2i\omega\tau}-\hat{a}^{2}e^{-2i\omega\tau}\right]$,
the matrix element $\langle0|\hat{K}^{I}(\tau)|1\rangle=0$. Then we just need to compute 
\begin{equation}
\langle0|\hat{K}^{I}(\tau)|0\rangle=\frac{\hbar\omega}{4}\langle0|\left[2\hat{N}+1-(\hat{a}^{\dagger})^{2}e^{+2i\omega\tau}-\hat{a}^{2}e^{-2i\omega\tau}\right]|0\rangle=\frac{\hbar\omega}{4}    
\end{equation}
and
\begin{equation}
\langle1|\hat{K}^{I}(\tau)|1\rangle=\frac{\hbar\omega}{4}\langle1|\left[2\hat{N}+1-(\hat{a}^{\dagger})^{2}e^{+2i\omega\tau}-\hat{a}^{2}e^{-2i\omega\tau}\right]|1\rangle=\frac{3\hbar\omega}{4},    
\end{equation}
from which we get $D_{2}(\tau)=\frac{3\hbar^{2}\omega^{2}}{32}$.
Putting all results together we have 
\begin{equation}
\langle0|e^{\mathcal{L}_{1}t}\left[\hat{K}^{I}(\tau),\left[\hat{K}^{I}(\tau),\hat{\rho}(0)\right]\right]|1\rangle\simeq e^{i\omega t}e^{-\frac{\gamma}{2}t}\left(\frac{3\hbar^{2}\omega^{2}}{32}-2\frac{3\hbar^{2}\omega^{2}}{32}+\frac{15\hbar^{2}\omega^{2}}{32}\right)=\frac{3\hbar^{2}\omega^{2}}{8}e^{i\omega t}e^{-\frac{\gamma}{2}t}\label{bobo2}
\end{equation}
Finally, taking the matrix elements $\langle0|...|1\rangle$ of Eq.
(\ref{App_sol_breu}) and using Eqs. (\ref{bobo1}) and (\ref{bobo2}) we get
\begin{equation}
\langle0|\hat{\rho}(t)|1\rangle\simeq\frac{1}{2}e^{i\omega t}e^{-\frac{\gamma}{2}t}\left(1-\frac{3}{8}\omega^{2}\tau_{c}t\right),\label{App_sol_breu-1}
\end{equation}
which is Eq.~\eqref{AAAAAAAAAA} of the main text.

\section{Study of the decay of the excited state}

Here we study the decay of the exited state, hence we take as initial
state $\hat{\rho}(0)=|1\rangle\langle1|$. Starting from Eq. (\ref{App_sol_breu})
and by taking the matrix element $\langle1|...|1\rangle$ one gets:
\begin{equation}
\langle1|\hat{\rho}(t)|1\rangle\simeq\langle1|e^{\mathcal{L}_{1}t}\hat{\rho}(0)|1\rangle-\frac{\tau_{c}}{2\hbar^{2}}\int_{0}^{t}d\tau\langle1|e^{\mathcal{L}_{1}t}\left[\hat{K}^{I}(\tau),\left[\hat{K}^{I}(\tau),\hat{\rho}(0)\right]\right]|1\rangle.\label{Sol_11_breu}
\end{equation}
The first term is the usual amplitude damping $\langle1|e^{\mathcal{L}_{1}t}\hat{\rho}(0)|1\rangle=e^{-\gamma t}$. The matrix element in the second term can be computed as
\begin{equation}
\langle1|e^{\mathcal{L}_{1}t}\left[\hat{K}^{I}(\tau),\left[\hat{K}^{I}(\tau),\hat{\rho}(0)\right]\right]|1\rangle\simeq e^{-\gamma t}\langle1|\left[\hat{K}^{I}(\tau),\left[\hat{K}^{I}(\tau),\hat{\rho}(0)\right]\right]|1\rangle
\end{equation}
with 
\begin{equation}\label{125}
\langle1|\left[\hat{K}^{I}(\tau),\left[\hat{K}^{I}(\tau),\hat{\rho}(0)\right]\right]|1\rangle=D_{1}^{(1)}(\tau)-2D_{2}^{(1)}(\tau)+D_{3}^{(1)}(\tau).
\end{equation}
The three terms in the RHS side of Eq. \eqref{125} are: 
\begin{align}
D_{1}^{(1)}(\tau)&:=\langle1|\hat{K}^{I2}(\tau)\hat{\rho}(0)|1\rangle=\langle1|\hat{K}^{I2}(\tau)|1\rangle=\frac{15\hbar^{2}\omega^{2}}{16};\label{D1_bre-1}
\\
D_{2}^{(1)}(\tau)&:=\langle1|\hat{K}^{I}(\tau)\hat{\rho}(0)\hat{K}^{I}(\tau)|1\rangle=\langle1|\hat{K}^{I}(\tau)|1\rangle\langle1|\hat{K}^{I}(\tau)|1\rangle=\left(\frac{3\hbar\omega}{4}\right)^{2}=\frac{9\hbar^{2}\omega^{2}}{16};\label{D2_bre-1}
\\
D_{3}^{(1)}(\tau)&:=\langle1|\hat{\rho}(0)\hat{K}^{I2}(\tau)|1\rangle=\langle1|\hat{K}^{I2}(\tau)|1\rangle=\frac{15\hbar^{2}\omega^{2}}{16}.\label{D3_bre-1}
\end{align}
Putting all these results together we get
\begin{equation}
\langle1|e^{\mathcal{L}_{1}t}\left[\hat{K}^{I}(\tau),\left[\hat{K}^{I}(\tau),\hat{\rho}(0)\right]\right]|1\rangle\simeq\frac{3\hbar^{2}\omega^{2}}{4}e^{-\gamma t},
\end{equation}
which substituted in Eq. (\ref{Sol_11_breu}) leads to:
\begin{equation}
\langle1|\hat{\rho}(t)|1\rangle\simeq e^{-\gamma t}\left(1-\frac{3}{8}\omega^{2}\tau_{c}t\right),
\end{equation}
which is Eq. \eqref{BBBBBBB} of the main text.

\section{Study of the evolution of the ground state}

We conclude by studying the evolution of the ground state, taking
as initial state $\hat{\rho}(0)=|0\rangle\langle0|$. Starting from
Eq. (\ref{App_sol_breu}) and by taking the matrix element $\langle0|...|0\rangle$
one gets:
\begin{equation}
\langle0|\hat{\rho}(t)|0\rangle\simeq\langle0|e^{\mathcal{L}_{1}t}\hat{\rho}(0)|0\rangle-\frac{\tau_{c}}{2\hbar^{2}}\int_{0}^{t}d\tau\langle0|e^{\mathcal{L}_{1}t}\left[\hat{K}^{I}(\tau),\left[\hat{K}^{I}(\tau),\hat{\rho}(0)\right]\right]|0\rangle.\label{Sol_11_breu-1}
\end{equation}
The first term it is simply $\langle0|e^{\mathcal{L}_{1}t}\hat{\rho}(0)|0\rangle=1$
while the matrix element in the second term can be computed as
\begin{equation}
\langle0|e^{\mathcal{L}_{1}t}\left[\hat{K}^{I}(\tau),\left[\hat{K}^{I}(\tau),\hat{\rho}(0)\right]\right]|0\rangle\simeq\langle0|\left[\hat{K}^{I}(\tau),\left[\hat{K}^{I}(\tau),\hat{\rho}(0)\right]\right]|0\rangle
\end{equation}
with 
\begin{equation}\label{133}
\langle0|\left[\hat{K}^{I}(\tau),\left[\hat{K}^{I}(\tau),\hat{\rho}(0)\right]\right]|0\rangle=D_{1}^{(0)}(\tau)-2D_{2}^{(0)}(\tau)+D_{3}^{(0)}(\tau).
\end{equation}
The three terms in the RHS side of Eq. \eqref{133} are: 
\begin{align}
D_{1}^{(0)}(\tau)&:=\langle0|\hat{K}^{I2}(\tau)\hat{\rho}(0)|0\rangle=\langle0|\hat{K}^{I2}(\tau)|0\rangle=\frac{3\hbar^{2}\omega^{2}}{16};\label{D1_bre-1-1}
\\
D_{2}^{(0)}(\tau)&:=\langle0|\hat{K}^{I}(\tau)\hat{\rho}(0)\hat{K}^{I}(\tau)|0\rangle=\langle0|\hat{K}^{I}(\tau)|0\rangle\langle0|\hat{K}^{I}(\tau)|0\rangle=\left(\frac{\hbar\omega}{4}\right)^{2}=\frac{\hbar^{2}\omega^{2}}{16};\label{D2_bre-1-1}
\\
D_{3}^{(0)}(\tau)&:=\langle0|\hat{\rho}(0)\hat{K}^{I2}(\tau)|0\rangle=\langle0|\hat{K}^{I2}(\tau)|0\rangle=\frac{3\hbar^{2}\omega^{2}}{16}.\label{D3_bre-1-1}
\end{align}
Putting all these results together we get
\begin{equation}
\langle0|e^{\mathcal{L}_{1}t}\left[\hat{K}^{I}(\tau),\left[\hat{K}^{I}(\tau),\hat{\rho}(0)\right]\right]|0\rangle\simeq\frac{\hbar^{2}\omega^{2}}{4},
\end{equation}
which substituted in Eq. (\ref{Sol_11_breu-1}) leads to:
\begin{equation}
\langle0|\hat{\rho}(t)|0\rangle\simeq1-\frac{\omega^{2}}{8}\tau_{c}t,
\end{equation}
which is Eq. \eqref{BBBBBBB} of the main text.

\end{widetext}

\end{document}